\documentclass{article}
\usepackage[utf8]{inputenc}
\usepackage{amsmath}
\usepackage{amssymb}
\usepackage{graphicx}
\usepackage{esint}
\usepackage{hyperref}
\hypersetup{
  colorlinks   = true,    
  urlcolor     = blue,    
  linkcolor    = blue,    
  citecolor    = black     
}

\newcommand\Tr{\mathrm{Tr}}
\newcommand*{\email}[1]{%
    \normalsize\href{mailto:#1}{#1}\par
    }

\usepackage{color}
\usepackage[makeroom]{cancel}
\usepackage[normalem]{ulem}
\definecolor{pkcolor}{rgb}{0,0.1,0.7}

\newcommand\pkout{\marginpar{\color{pkcolor}$\clubsuit$}\bgroup\markoverwith{\color{pkcolor}{\rule[0.4ex]{2pt}{0.8pt}}}\ULon}

\definecolor{hkcolor}{rgb}{0.7,0.0,0.0}

\newcommand\hkout{\marginpar{\color{hkcolor}$\clubsuit$}\bgroup\markoverwith{\color{hkcolor}{\rule[0.4ex]{2pt}{0.8pt}}}\ULon}

\definecolor{ascolor}{rgb}{0.1,0.7,0.0}

\DeclareRobustCommand\asout{\bgroup\markoverwith{\color{ascolor}{\rule[0.4ex]{2pt}{0.8pt}}}\ULon}

\title{Exploring straight infinite Wilson lines in the Self Dual and the MHV Lagrangians}
\author{Hiren Kakkad\footnote{\email{kakkad@agh.edu.pl}}$\,\,\,^a$,
Piotr Kotko\footnote{\email{pkotko@agh.edu.pl}}$\,\,\,^a$,
Anna Stasto\footnote{\email{ams52@psu.edu}}$\,\,\,^b$
\\ \\
$^a${\it AGH University Of Science and Technology, Physics Faculty,} \\ 
{\it Mickiewicza 30, 30-059 Krakow, Poland} \\ \\
$^b${\it The Pennsylvania State University, Physics Department}\\ 
{\it 104 Davey Lab, University Park, PA 16802, USA }
}

\date{}

\begin{document}
\maketitle

\begin{abstract}
    We investigate the appearance of straight infinite Wilson lines lying 
    on the self-dual plane in the context of the Self Dual sector
    of the Yang Mills theory and in a connection to the Lagrangian
    implementing the MHV vertices (MHV Lagrangian) according to the Cachazo-Svrcek-Witten method.
    It was already recognized in the past by two of the authors, that such Wilson
    line functional provides the field transformation of positive helicity fields
    between the Yang-Mills theory on the light-cone and the MHV Lagrangian.
    Here we discuss in detail the connection to the Self Dual sector and we provide
    a new insight into the solution for the minus helicity field transformation,
    which can be expressed in terms of a functional derivative of the straight infinite Wilson line on the self-dual plane.
\end{abstract}

\section{Introduction}
\label{sec:Intro}

Since the publication of the famous paper on Maximally Helicity Violating (MHV) amplitudes by Tomasz Taylor and Stephen Parke \cite{Parke:1986gb}, there has been an enormous development in methods of calculating scattering amplitudes in Quantum Chromodynamics (QCD), \emph{without use} of the Feynman diagrams. 
 Indeed, this direct technique, which is more than a half of century old, is extremely inefficient when confronted with today's needs. Proliferation of diagrams is huge even for small multiplicity processes. Moreover, individual diagrams do not respect gauge invariance leading to large cancellations between distinct sets of diagrams. In addition, other more subtle properties of the theory are not utilized. In particular, the incredible simplicity of the Parke-Taylor result for the tree-level MHV amplitudes has been recognized as a result of a `geometry' of the theory. It turns out, that  these amplitudes are supported on a line in the twistor space (four dimensional complex projective space) \cite{Witten2004}, which in turn corresponds to a point in Minkowski space. Therefore, the MHV amplitudes constitute `interaction vertices' in the Minkowski space. Indeed,  arbitrary tree amplitude can be constructed by gluing together  MHV amplitudes, what is now known as the Cachazo-Svrcek-Witten (CSW) method \cite{Cachazo2004}.
 This lead further to discovery of on-shell recursion relations \cite{Britto:2004ap,Britto:2005fq}, the so-called Britto-Cachazo-Feng-Witten (BCFW) method. It allows for constructing an on-shell amplitude with arbitrary number of external legs, from the on-shell amplitudes with less number of legs where the external momenta have been deformed to the complex domain. Since the building blocks are on-shell, the recursion is fully gauge invariant. It was also shown that the CSW method can be viewed as a particular case of the BCFW recursion \cite{Risager2005}. 
 In parallel, the on-shell conditions for internal lines were also intensively used to progress in calculation of loop amplitudes and are known as unitarity methods \cite{Bern1994,Bern2007,Brandhuber2008,Perkins2009,Bern2011}. 
 In recent years the on-shell methods have been also intensively studied from a completely geometric point of view. 
 In particular it was recognized that the external data describing scattering amplitudes (i.e. the momenta in spinor representation together with momentum conservation) constitute so-called positive Grassmannian (see  \cite{Arkani-Hamed_book_2016} for a pedagogical review and original references). This lead to the concept of the so-called amplituhedron \cite{Arkani-Hamed2014} whose volume gives scattering amplitudes in maximally supersymmetric theory. This concept has been recently extended also to non-supersymmetric theories \cite{
Arkani-Hamed2018a}.

 Let us point out, that besides exploring on-shell methods it is also important to study full off mass shell dynamics of fields. Although for some theories the scattering matrix (targeted in on-shell methods) can be considered as a physical object, in QCD it is not the case due to the color confinement. Therefore, one needs a bridge connecting physical hadrons with partons. This is usually achieved by means of collinear factorization theorems, which utilize the on-shell scattering amplitudes for the short distance part of the process and parton distribution functions and fragmentation functions for the non-perturbative part \cite{Collins:1985ue}. They depend however on the kinematics of a scattering process and not all processes can be factorized into non-perturbative hadronic wave function and  on-shell scattering amplitude. Notably, there is a large class of processes occurring at high energies that are described in terms of so-called $k_T$-factorization \cite{Catani:1990eg,Collins1991,Catani:1994sq}, which utilizes gauge invariant \emph{off-shell} amplitudes (equivalent to Reggeon scattering amplitudes in high-energy kinematics) \cite{Antonov:2004hh,VanHameren2012, VanHameren2013a,Kotko2014b,vanHameren:2014iua}.
 Therefore, it is also important to study off-mass shell correlators and form factors, beyond purely on-shell functions. Application of geometric methods has been less intensively studied in that context, but some interesting results do exist \cite{Bork2015,Frassek2016,Bork2017}.
 Due to the above, study of dynamics of fields in more traditional way, i.e. using quantum field theory, is also important.

 In the present work, we continue exploring the connection of the MHV vertices used in the CSW method and straight infinite Wilson lines on certain complex plane, that has been started in \cite{Kotko2017}.
 These Wilson lines emerge as the transformation of the positive helicity Yang-Mills field appearing in the light cone action, to a new action where the MHV vertices are explicit \cite{Mansfield2006,Ettle2006b,Ettle2007,Ettle2008}. 
As a first objective we explore  the connection of this Wilson line to the Self Dual sector of the Yang-Mills theory. The scattering amplitudes in that sector have been discussed in detail in the literature \cite{Bardeen1996,Chalmers1996,Cangemi1997,Rosly1997,Monteiro2011}, as well as the integrability of the self dual equations of motion  \cite{Yang1977,Ward1977a,Ward1990,Mason1993}. 
Despite the fact that there exists rich and mathematically sophisticated  literature on the subject, there is no direct discussion of the infinite Wilson line discovered in \cite{Kotko2017} in that context.
The second objective of the paper is to extend the discussion beyond the self dual sector. As mentioned, the straight infinite Wilson line functional considered in \cite{Kotko2017} transforms the positive helicity fields to new fields that constitute the MHV action. Here, we present an analogous solution for the minus helicity field, which turns out to be transformed by certain functional derivative of the Wilson line.

The structure of the paper is the following. In Section~\ref{sec:SD} we discuss the self dual sector. We recall the equations of motion and structure of the tree level currents existing in the theory. Next, we discuss how the straight infinite Wilson line emerges as an inverse functional to the generating function for the solutions. In Section~\ref{sec:MHV} we extend the discussion of Wilson lines beyond the Self Dual sector. We find a direct relation of the Wilson line from the self dual sector to the transformation of the minus helicity fields that leads to the MHV action. The last Section is devoted to a summary and description of future prospects.

\section{The Self Dual Yang-Mills Theory}
\label{sec:SD}

\subsection{Classical EOM and scattering amplitudes} 
\label{sub:SD_EOM}

Our starting point is the full Yang-Mills action in the light-cone gauge $A^+=0$, where the $A^-$ fields (appearing quadratically) were integrated out from the theory \cite{Scherk1975} leaving only two  complex fields $A^{\bullet}$, $A^{\star}$ that correspond to plus-helicity and minus-helicity gluon fields. We use here very convenient so-called 'double-null' coordinates defined as 
$v^{+}=v\cdot\eta$, $v^{-}=v\cdot\tilde{\eta}$, 
$v^{\bullet}=v\cdot\varepsilon_{\bot}^{+}$,  
$v^{\star}=v\cdot\varepsilon_{\bot}^{-}$
with the two light-like basis four-vectors 
$\eta=\left(1,0,0,-1\right)/\sqrt{2}$, $\tilde{\eta}=\left(1,0,0,1\right)/\sqrt{2}$,
and two space like complex four-vectors spanning the transverse plane
$\varepsilon_{\perp}^{\pm}=\frac{1}{\sqrt{2}}\left(0,1,\pm i,0\right)$. With the above definitions one can easily  raise and lower indices, taking care of the flip $+\leftrightarrow -$ and $\star \leftrightarrow  \bullet$, where the latter causes also the sign change. 
The action can be written as 
\begin{equation}
S_{\mathrm{Y-M}}^{\left(\mathrm{LC}\right)}\left[A^{\bullet},A^{\star}\right]=\int dx^{+}\left(\mathcal{L}_{+-}^{\left(\mathrm{LC}\right)}+\mathcal{L}_{++-}^{\left(\mathrm{LC}\right)}+\mathcal{L}_{+--}^{\left(\mathrm{LC}\right)}+\mathcal{L}_{++--}^{\left(\mathrm{LC}\right)}\right)\,,\label{eq:actionLC}
\end{equation}
where the individual terms in the Lagrangian read:
\begin{gather}
\mathcal{L}_{+-}^{\left(\mathrm{LC}\right)}\left[A^{\bullet},A^{\star}\right]=-\int d^{3}\mathbf{x}\,\mathrm{Tr}\,\hat{A}^{\bullet}\square\hat{A}^{\star}\,,\\
\mathcal{L}_{++-}^{\left(\mathrm{LC}\right)}\left[A^{\bullet},A^{\star}\right]=-2ig\,\int d^{3}\mathbf{x}\,\mathrm{Tr}\,\partial_{-}^{-1}\partial_{\bullet} \hat{A}^{\bullet}\left[\partial_{-}\hat{A}^{\star},\hat{A}^{\bullet}\right]\,,\\
\mathcal{L}_{--+}^{\left(\mathrm{LC}\right)}\left[A^{\bullet},A^{\star}\right]=-2ig\,\int d^{3}\mathbf{x}\,\mathrm{Tr}\,\partial_{-}^{-1}\partial_{\star}\hat{A}^{\star}\left[\partial_{-}\hat{A}^{\bullet},\hat{A}^{\star}\right]\,,\\
\mathcal{L}_{++--}^{\left(\mathrm{LC}\right)}\left[A^{\bullet},A^{\star}\right]=-2g^{2}\int d^{3}\mathbf{x}\,\mathrm{Tr}\,\left[\partial_{-}\hat{A}^{\bullet},\hat{A}^{\star}\right]\partial_{-}^{-2}\left[\partial_{-}\hat{A}^{\star},\hat{A}^{\bullet}\right]\, ,
\end{gather}
where $\square=2(\partial_+\partial_- - \partial_{\bullet}\partial_{\star})$. 
Above, the bold position vector is defined as $\mathbf{x}\equiv\left(x^{-},x^{\bullet},x^{\star}\right)$. We denote $\hat{A}=A_at^a$ as the color algebra element and use the normalization of color generators that fulfills $\left[t^{a},t^{b}\right]=i\sqrt{2}f^{abc}t^{c}$. With this normalization, our coupling constant is re-scaled as $g\rightarrow g/\sqrt{2}$ comparing to 'standard' normalization of the action.

The fully covariant form of the self dual equations is
\begin{equation}
    \hat{F}^{\mu\nu} = \ast \hat{F}^{\mu\nu}\, ,
\end{equation}
where 
$\hat{F}^{\mu\nu}=\partial^{\mu}\hat{A}^{\nu}-\partial^{\nu}\hat{A}^{\mu} - ig \left[ \hat{A}^{\mu},\hat{A}^{\nu} \right]$ and the Hodge dual is defined as 
$\ast \hat{F}_{\mu\nu} = -i \epsilon_{\mu\nu\alpha\beta}\hat{F}^{\alpha\beta}$.
The corresponding self dual equation in light-cone gauge and upon eliminating the $A^-$ field reads 
\begin{equation}
    \Box {\hat{A}}^{\bullet} + 2ig{\partial}_{-} \left[ ({\partial}_{-}^{-1} {\partial}_{\bullet} {\hat{A}}^{\bullet}), {\hat{A}}^{\bullet} \right] = 0\, ,
    \label{eq:SD_EOM}
\end{equation}
which can be obtained from the following truncation of the full action 
\begin{equation}
    S_{\mathrm{SDYM}}^{(\mathrm{LC})} [A^{\bullet}, A^{\star}] = \int\! dx^{+} \left[ \mathcal{L}^{(\mathrm{LC})}_{+-} +\mathcal{L}^{(\mathrm{LC})}_{++-} \right]\, .
    \label{eq:actionSD}
\end{equation}

\begin{figure}
    \centering
    \includegraphics[width=8cm]{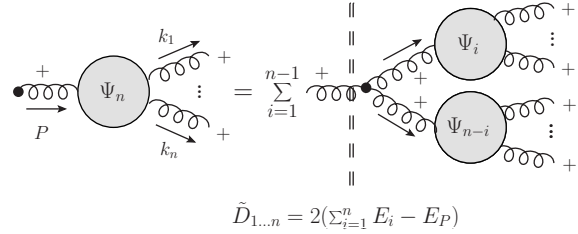}
    \caption{
    \small
    The tree-level off-shell currents generated by the classical solution to the self-dual theory satisfy the light-cone variant of the Berends-Giele recursion. The plus signs indicate the helicity of gluons. The vertical double-dashed line denotes the energy denominator $\tilde{D}_{1\dots n}=2(\sum_{i=1}^n E_i - E_P)$, where the light-cone energy for a momentum $k$ is $E_k=k^{\bullet}k^{\star}/k^+$. } 
    \label{fig:AbulletBG}
\end{figure}

We are interested in the classical solution to the self-dual EOM (\ref{eq:SD_EOM}) that give us information about scattering amplitudes.
The tree-level Green functions can be extracted by coupling the classical action to an external current
\begin{equation}
W[A^{\bullet}, A^{\star};J]= S_{\mathrm{SDYM}}^{(\mathrm{LC})} [A^{\bullet}, A^{\star}]+\int\! dx^+\, \mathrm{Tr}\hat{J}\hat{A}^{\star}
\label{eq:SDaction_J}
\end{equation}
and postulating the power series solution:
\begin{equation}
    A_a^{\bullet}\left[j\right](x) = \sum_{n=1}^{\infty} 
    \int d^4y_1\dots d^4y_n \, \Psi_n^{ab_1\dots b_n}(x;y_1,\dots ,y_n) j_{b_1}(y_1)\dots j_{b_n}(y_n)\,.
    \label{eq:Abullet_sol1}
\end{equation}
Above, the current $J$ has been replaced by a current that accommodates the on-shell pole 
\begin{equation}
    j_a(x)= \square^{-1}  J_a(x) \, .
    \label{eq:j_def}
\end{equation} 
Assuming the currents $j_a$ are supported on the light-cone, the  
momentum space off-shell currents $\tilde{\Psi}^{ab_1\dots b_n}$  generated by the solution (\ref{eq:Abullet_sol1}) correspond to the off-shell currents similar to the Berends-Giele currents \cite{Berends:1987me}, i.e. to  an amplitude of an off-shell gluon with momentum $P$ and plus helicty (momentum incoming) scattering into $n$ on-shell gluons with momenta $k_1,\dots, k_n$. The solution can be obtained iteratively, see Fig.~\ref{fig:AbulletBG}
and reads \cite{Cangemi1997,Motyka2009}
\begin{multline}
    \tilde{\Psi}_n^{a\{b_1\dots b_n\}}\left(P;\{p_1,\dots ,p_n\}\right) =     
    - (-g)^{n-1}  
    \delta^{4} (p_{1} + \cdots + p_{n} - P) \\
    \Tr \left(t^{a} t^{b_{1}} \cdots t^{b_{n}}\right) \,  
    \frac{{\tilde v}^{\ast}_{(1 \cdots n)1}}{{\tilde v}^{\ast}_{1(1 \cdots n)}}
    \frac{1}{{\tilde v}^{\ast}_{21}{\tilde v}^{\ast}_{32} \cdots {\tilde v}^{\ast}_{n(n-1)}} \, ,
    \label{eq:Psi_n}
\end{multline}
where we introduced spinor-like variables following \cite{Cruz-Santiago2015}
\begin{equation}
    \tilde{v}_{ij}=
    p_i^+\left(\frac{p_{j}^{\star}}{p_{j}^{+}}-\frac{p_{i}^{\star}}{p_{i}^{+}}\right), \qquad 
\tilde{v}^*_{ij}=
    p_i^+\left(\frac{p_{j}^{\bullet}}{p_{j}^{+}}-\frac{p_{i}^{\bullet}}{p_{i}^{+}}\right)\, .
\label{eq:vtilde}
\end{equation}
We use a shorthand notation for the sum of momenta $p_1+\dots p_n\equiv p_{1\dots n}$. The curly braces denote the symmetrization with respect to color and momentum variables, i.e. pairs $\{b_i,p_i\}$.
Here and in what follows we always skip factors of $2\pi$ accompanying conservation delta functions.
    The $\tilde{v}_{ij}$, $\tilde{v}_{ij}^*$ symbols have the property 
that  in addition to being directly proportional to the spinor products
$\left< ij \right>$ and $\left[ ij \right]$ (see. e.g.
\cite{Cruz-Santiago2015}), they can be defined in terms of the polarization vectors
\begin{equation}
    \varepsilon^{\pm}_i=\varepsilon_{\perp}^{\pm}-\frac{p_i\cdot \varepsilon_{\perp}^{\pm}}{p_i\cdot\eta}\, \eta \; ,
    \label{eq:polvec}
\end{equation}
as follows
\begin{equation}
    \tilde{v}^*_{ij}=-(\varepsilon_i^+\cdot k_j)\,, \quad 
    \tilde{v}_{ij}=-(\varepsilon_i^-\cdot k_j )\, . 
    \label{eq:vtilde1}
\end{equation}
This turns out to be important in the context of the inverse functional to (\ref{eq:Abullet_sol1}) as we discuss in the next subsection.

The off-shell currents $\tilde{\Psi}_n^{ab_1\dots b_n}$ can be turned into on-shell $-,+,\dots,+$ scattering amplitude by putting the incoming off-shell gluon on mass shell and amputating the propagator, or equivalently, the energy denominator 
\begin{equation}
    \tilde{D}_{1\dots n} = 2\left(\sum_{i=1}^n E_i - E_{(1\dots n)}\right) \, .
    \label{eq:Dtilde}
\end{equation}
Above, the the light-cone energy 
\begin{equation}
    E_i = \frac{k_i^{\bullet}k_i^{\star}}{k_i^+} \,  ,
    \label{eq:LCenergy}
\end{equation}
is equal to the 'minus' momentum component obtained from the on-shell condition $k_i^2=0$.
Since the solution (\ref{eq:Psi_n}) does not have the $1/\tilde{D}_{1\dots n}$ factor (it gets canceled -- see Fig.~\ref{fig:AbulletBG})  we get zero upon amputation and imposing the on-shell limit, i.e. when the minus component is conserved. 
Thus the on-shell $-,+,\dots,+$ tree amplitude is zero in agreement with well known general analysis.

\subsection{The straight infinite Wilson line}
\label{sub:SD_WL}

It is very interesting how the Self Dual equation encodes the infinite Wilson line spanning over the transverse complex plane. Obviously, due to the integrability property of the theory the relation to Wilson lines should not be a surprise \cite{Yang1977}, but the following exposition is new and quite elementary.

Let us consider an inverse functional to (\ref{eq:Abullet_sol1}). First note, that the kernels $\tilde{\Psi}_n$ are functions exclusively of the three momenta $\mathbf{p}\equiv\left(p^{+},p^{\bullet},p^{\star}\right)$, up to the `minus' component momentum conservation delta function $\delta(p_1^- + \dots + p_n^- - P^-)$. Therefore 
we define the 3D kernels $\tilde{\Psi}_n^{a\{b_1\dots b_n\}}\left(\mathbf{P};\{\mathbf{p}_1,\dots ,\mathbf{p}_n\}\right)$ using
\begin{multline}
    \tilde{\Psi}_n^{a\{b_1\dots b_n\}}\left(P;\{p_1,\dots ,p_n\}\right) = 
    \delta\left(p_1^- + \dots + p_n^- - P^-\right)
    \tilde{\Psi}_n^{a\{b_1\dots b_n\}}\left(\mathbf{P};\{\mathbf{p}_1,\dots ,\mathbf{p}_n\}\right) \, ,
    \label{eq:Psi_n_3D}
\end{multline}
We can get rid of the 'minus' component delta function in (\ref{eq:Abullet_sol1}) by Fourier transforming it with respect to the 'minus' momentum component:
\begin{multline}
    \tilde{A}_a^{\bullet}\left[j\right](x^+;\mathbf{P}) = \sum_{n=1}^{\infty} 
    \int d^3\mathbf{p}_1\dots d^3\mathbf{p}_n \, \tilde{\Psi}_n^{a\{b_1\dots b_n\}}(\mathbf{P};\{\mathbf{p}_1,\dots ,\mathbf{p}_n\}) \\ \tilde{j}_{b_1}(x^+;\mathbf{p}_1)\dots \tilde{j}_{b_n}(x^+;\mathbf{p}_n)\,.
    \label{eq:Abullet_sol2}
\end{multline}
We assume that the inverse functional has the expansion
\begin{multline}
    \tilde{j}_a\left[A^{\bullet}\right](x^+;\mathbf{P}) = \sum_{n=1}^{\infty} 
    \int d^3\mathbf{p}_1\dots d^3\mathbf{p}_n \, \tilde{\Gamma}_n^{a\{b_1\dots b_n\}}(\mathbf{P};\{\mathbf{p}_1,\dots ,\mathbf{p}_n\}) \\ \tilde{A^{\bullet}}_{b_1}(x^+;\mathbf{p}_1)\dots \tilde{A^{\bullet}}_{b_n}(x^+;\mathbf{p}_n)\,.
    \label{eq:j[A]}
\end{multline}
Inserting (\ref{eq:j[A]}) to (\ref{eq:Abullet_sol2}) one finds 
\footnote{
Virtually identical calculation has been done in \cite{Kotko2017} in the context of a transformation between Yang-Mill action and the MHV action. 
In this section we repeat these steps referring solely to the Self Dual action.}
\begin{equation}
    \tilde{\Gamma}^{a\{b_1\dots b_n\}}_n (\mathbf{P};\{\mathbf{p}_1,\dots,\mathbf{p}_n\}) = (-g)^{n-1} \frac{\delta^3\left(\mathbf{p}_1+\dots+\mathbf{p}_n-\mathbf{P}\right)\,\Tr\!\left(t^at^{b_1}\dots t^{b_n}\right)}{\Tilde{v}^*_{1(1\cdots n)} \Tilde{v}^*_{(12)(1\cdots n)} \cdots \Tilde{v}^*_{(1 \cdots n-1)(1\cdots n)}} \, .
    \label{eq:Gamma_n}
\end{equation}
Noticing that each of the factors in the denominator is actually a scalar product of a sum of subsequent momenta and the \emph{same} polarization vector (\ref{eq:polvec}): 
\begin{equation}
    \tilde{v}^*_{(1\dots i)(1\dots n)} = -
    \left(p_1 + \dots + p_i \right) \cdot \varepsilon^+_P \, ,
\end{equation}
it is not difficult to prove \cite{Kotko2017} 
that (\ref{eq:Gamma_n}) are the momentum space expansion coefficients of the functional directly related to the following straight infinite Wilson line along $\varepsilon^+_P$ (in the light-cone gauge): 
\begin{equation}
j_{a}[A]\left(x\right)=\int_{-\infty}^{\infty}d\alpha\,\mathrm{Tr}\left\{ \frac{1}{2\pi g}t^{a}\partial_{-}\, \mathbb{P}\exp\left[ig\int_{-\infty}^{\infty}ds\, \varepsilon_{\alpha}^+\cdot \hat{A}\left(x+s\varepsilon_{\alpha}^+\right)\right]\right\} \, ,
\label{eq:WilsonLineSol}
\end{equation}
where 
\begin{equation}
    \varepsilon_{\alpha}^{\pm} = \epsilon_{\perp}^{\pm}- \alpha \eta \, .
    \label{eq:epsilon_alpha}
\end{equation}
is a generic polarization vector. 
Thus, from the position space point of view the functional (\ref{eq:WilsonLineSol}) is the (derivative of) Wilson line integrated over all possible directions. However, after 
 passing to momentum space and integrating over $\alpha$, the generic polarization vector (\ref{eq:epsilon_alpha}) becomes $\varepsilon_P^+$.
 
 One may worry that the functional (\ref{eq:WilsonLineSol}) is in contradiction with the EOM with the current (compare \eqref{eq:SD_EOM}): 
\begin{equation}
    \Box {\hat{A}}^{\bullet} + 2ig{\partial}_{-} \left[ ({\partial}_{-}^{-1} {\partial}_{\bullet} {\hat{A}}^{\bullet}), {\hat{A}}^{\bullet} \right] - \Box\, \hat{j} = 0\, ,
    \label{eq:SD_EOM_j}
\end{equation}
as it contains terms at most quadratic in the fields, whereas (\ref{eq:WilsonLineSol}) contains all powers. However, the current $j$ in (\ref{eq:SD_EOM_j}) is supported on the light-cone and only the second term in the expansion of (\ref{eq:WilsonLineSol}), $\tilde{\Gamma}_2$, contains implicit pole $1/P^2$. We discuss this point in Appendix~\ref{sec:AppA}. 
In view of the above, the functional (\ref{eq:WilsonLineSol}) is a more general object then the external current $j$ and therefore we rename it as follows:
\begin{equation}
    B^{\bullet}_a[A](x) \equiv j_{a}[A]\left(x\right)\, .
    \label{eq:B_field}
\end{equation}

The Wilson line (\ref{eq:WilsonLineSol}) has been already recognized in \cite{Kotko2017} as the relation between the positive helicity Yang-Mills field and positive helicity field in the MHV action -- we shall come back to this point in the next section.  It is also worth mentioning  that interestingly the kernels $\tilde{\Gamma}_n$ contain the same diagrams as $\tilde{\Psi}_n$, but different energy denominators, see \cite{Kotko2017} for details.  

Let us now discuss the origin of the Wilson line (\ref{eq:WilsonLineSol})  in more geometrical terms. 

Notice first, that this Wilson line lies within the  plane spanned by $\eta$ and $\epsilon_{\perp}^+$.  It is a null plane, i.e. any tangent vector is light-like. It is  interesting, that this plane is also a self-dual plane, i.e.  any bivector $w^{\mu\nu}=u^{\mu}v^{\nu}-u^{\nu}v^{\mu}$ with tangent vectors $u$,$v$ is self-dual (such plane is also called $\alpha$-plane or $\beta$-plane in the twistor space formulation \cite{Ward1977a}). Therefore, any Wilson line lying within the plane is the functional of the self-dual connection $A$ and one can attempt to reconstruct the connection from the Wilson line.

\begin{figure}
    \centering
    \includegraphics[width=8cm]{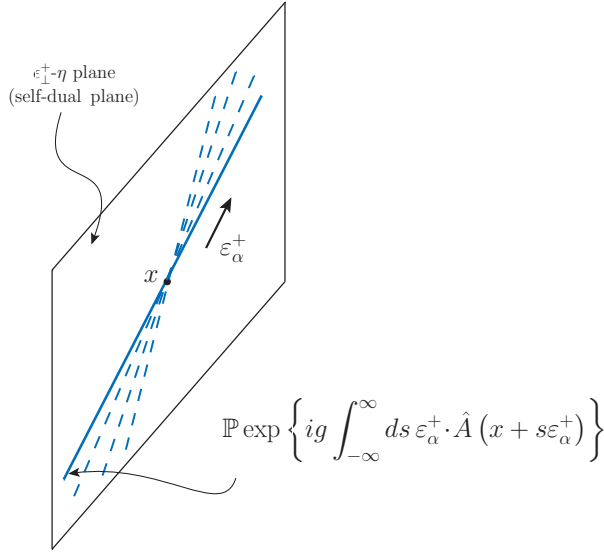}
    \caption{
    \small
    The inverse functional to the generating functional for the self dual solutions is given by the straight infinite Wilson line lying on the plane spanned by $\varepsilon^+_{\alpha}=\epsilon^+_{\perp}-\alpha\eta$  (with $\epsilon^+_{\perp}=(0,1,i,0)/\sqrt{2}$, $\eta=(1,0,0,-1)/\sqrt{2}$) and integrated over all $\alpha$ (the dashed lines represent tilted Wilson lines due to the change of $\alpha$).
    } 
    \label{fig:SDplane}
\end{figure}

A generic Wilson line functional $U_C[A](x,y)$ depends on two endpoints and a path $C$, and thus cannot be used to solve for the self-dual gauge field. 
The trick is now, to choose the simplest possible path crossing arbitrary point and extending over the whole plane, i.e. the straight line (see Fig.~\ref{fig:SDplane}). Let us assume we pick up the line along the vector $\varepsilon_{\alpha}$ defined in (\ref{eq:epsilon_alpha}). In addition to being independent on the end-points, the resulting infinite Wilson line $U_{\alpha}[A](x)$ is gauge invariant with respect to the small gauge transformations. Finally, integration over $\alpha$ reduces the functional $U_{\alpha}[A](x)$ to a functional independent on the path details $U[A](x)$. One can now find an inverse functional $A[U](x)$ which is the generating functional for the off-shell amplitudes.
Notice the functional $U$ can be treated here as the fundamental representation field
as it depends only on color and position. It is the origin of the $B^{\bullet}$ field defined in (\ref{eq:B_field}).

Although the procedure outlined above is quite intuitive, the detailed mathematical treatment and connection to integrability of the self-dual sector is beyond the scope of the present paper.

\section{The MHV Lagrangian}
\label{sec:MHV}

In the previous section we recalled the self dual Yang-Mills theory and  showed how straight infinite Wilson line functional, integrated over all directions, emerges as an inverse functional to the solution of the EOM. It can be interpreted as a new field $B^{\bullet}_a$ in the fundamental color representation, see Eq.~(\ref{eq:B_field}).

It turns out \cite{Kotko2017} that mapping $A^{\bullet}\rightarrow B^{\bullet}[A^{\bullet}]$ given by  (\ref{eq:WilsonLineSol}) is one of the two Mansfield transformations taking the full Yang-Mills action to the action with MHV vertices \cite{Mansfield2006}. The second transformation relates the starred fields $A^{\star}\rightarrow B^{\star}[A^{\bullet},A^{\star}]$ and will be discussed in more details below. 

\subsection{Recap of the MHV action}
\label{sub:MHVaction}

The action involving MHV vertices reads \cite{Mansfield2006}:
\begin{equation}
S_{\mathrm{Y-M}}^{\left(\mathrm{LC}\right)}\left[\tilde{B}^{\bullet},\tilde{B}^{\star}\right]=\int dx^{+}\left(\mathcal{L}_{+-}^{\left(\mathrm{LC}\right)}+\mathcal{L}_{--+}^{\left(\mathrm{LC}\right)}+\dots+\mathcal{L}_{--+\dots+}^{\left(\mathrm{LC}\right)}+\dots\right)\,,\label{eq:MHV_action}
\end{equation}
where the form of the kinetic term is 
analogous to the one in (\ref{eq:actionLC}) 
but with $A$ fields replaced by the $B$ fields

whereas the $n$-point interaction terms are
\begin{multline}
\mathcal{L}_{--+\dots+}^{\left(\mathrm{LC}\right)}=\int d^{3}\mathbf{p}_{1}\dots d^{3}\mathbf{p}_{n}\delta^{3}\left(\mathbf{p}_{1}+\dots+\mathbf{p}_{n}\right)\,
\tilde{\mathcal{V}}_{--+\dots+}^{b_{1}\dots b_{n}}\left(\mathbf{p}_{1},\dots,\mathbf{p}_{n}\right)
\\ \tilde{B}_{b_{1}}^{\star}\left(x^+;\mathbf{p}_{1}\right)\tilde{B}_{b_{2}}^{\star}\left(x^+;\mathbf{p}_{2}\right)\tilde{B}_{b_{3}}^{\bullet}\left(x^+;\mathbf{p}_{3}\right)\dots\tilde{B}_{b_{n}}^{\bullet}\left(x^+;\mathbf{p}_{n}\right)
\,,
\label{eq:MHV_n_point}
\end{multline}
with the MHV vertices
\begin{multline}
\tilde{\mathcal{V}}_{--+\dots+}^{b_{1}\dots b_{n}}\left(\mathbf{p}_{1},\dots,\mathbf{p}_{n}\right)=
g^{n-1}\left(\frac{p_{1}^{+}}{p_{2}^{+}}\right)^{2}
\mathrm{Tr}\left(t^{b_1}\dots t^{b_n}\right) \\
\frac{\tilde{v}_{21}^{*4}}{\tilde{v}_{1n}^{*}\tilde{v}_{n\left(n-1\right)}^{*}\tilde{v}_{\left(n-1\right)\left(n-2\right)}^{*}\dots\tilde{v}_{21}^{*}}
\,.
\label{eq:MHV_vertex}
\end{multline}

The Wilson line functional (\ref{eq:WilsonLineSol}) appears here naturally as a consequence of mapping the $-++$ helicity vertex  of the Yang-Mills theory, which is not present in the CSW method, to a new free theory of $B$ fields, as proposed in \cite{Mansfield2006}. This vertex constitutes the self-dual theory as discussed in the previous section.

Because there are at most two $B^{\star}$ fields for arbitrarily large vertex, we see that the MHV action is almost entirely occupied by the Wilson lines. Of course it is the 'impurity' injected by the $B^{\star}$ fields that gives the rich structure of the full Yang-Mills theory.

\subsection{Relation of $B^{\star}$ field to infinite Wilson line}
\label{sub:Bstar}

The structure of the $B^{\star}$ field has not been discussed in the literature in a more geometrical manner. Actually, only the functional $A^{\star}[B^{\bullet},B^{\star}]$ 
 and not its inverse 
has been so far calculated \cite{Ettle2006b}, as this is what is needed to derive the MHV action.
Here we are interested in exploring the structure of the new field $B^{\star}$  as seen in the original theory.

In order to find an expression for $B^{\star}[A^{\bullet},A^{\star}]$ similar to (\ref{eq:WilsonLineSol}), we first derive the functional in momentum space, as a power series. This is done in a similar way as for the $B^{\bullet}$ field (see Appendix~\ref{sec:AppB} for details). We obtain
\begin{multline}
    {\tilde{B}}_a^\star (x^+;\mathbf{P}) = {\tilde A}^\star_{a} (x^+;\mathbf{P}) + \sum_{n=2}^{\infty} \int\!d^{3}\mathbf{p_1}\cdots d^{3}\mathbf{p_n} {\tilde \Upsilon}_{n}^{a b_1 \left \{b_2 \cdots b_n \right \} }(\mathbf{P}; \mathbf{p_1} ,\left \{ \mathbf{p_2} , \cdots \mathbf{p_n} \right \})  \\
     {\tilde A}^\star_{b_1} (x^+;\mathbf{p_1}){\tilde A}^\bullet_{b_2} (x^+;\mathbf{p_2}) \cdots {\tilde A}^\bullet_{b_n} (x^+;\mathbf{p_n}) \, ,
   \label{eq:Bstar}
\end{multline}
where
\begin{equation}
    {\tilde \Upsilon}_{n}^{a b_1 \left \{b_2 \cdots b_n \right \} }(\mathbf{P}; \mathbf{p_1} ,\left \{ \mathbf{p_2} , \cdots \mathbf{p_n} \right \}) = n\left(\frac{p_1^+}{p_{1\cdots n}^+}\right )^2 {\tilde \Gamma}_{n}^{a b_1 \cdots b_n }(\mathbf{P}; \mathbf{p_{1}}  \cdots \mathbf{p_{n}} ) \, .
    \label{eq:Upsilon_n}
\end{equation}

Note, that the kernels (\ref{eq:Upsilon_n}) are very similar to the Wilson line kernels (\ref{eq:Gamma_n}), modulo the prefactor. Indeed, it turns out that the solution for the $B^{\star}$ field has the following compact functional form in the position space:
\begin{multline}
    B_a^{\star}(x) = 
    \int_{-\infty}^{+\infty}\! d\alpha\,\, 
    \mathrm{Tr} \Big\{
    \frac{1}{2\pi g} t^a \partial_-^{-1} 
    \int\! d^4y \,
     \left[\partial_-^2 {A}_c^{\star}(y)\right]  \\
     \frac{\delta}{\delta {A}_c^{\bullet}(y)} \,
    \mathbb{P} \exp {\left[ig \int_{-\infty}^{+ \infty}\! ds \,
    \hat{A}^{\bullet}(x+s\varepsilon_{\alpha})\right]  } 
    \Big\} \,.
    \label{eq:Bstar_WL}
\end{multline}
The proof of this formula is given in Appendix~\ref{sec:AppB}.
Because the infinite Wilson line in (\ref{eq:Bstar_WL}) is the same as in (\ref{eq:WilsonLineSol}) we can write the above formula in terms of the  functional derivative of $B^{\bullet}$ field:
\begin{equation}
    B_a^{\star}(x) = 
    \int\! d^3\mathbf{y} \,
     \left[ \frac{\partial^2_-(y)}{\partial^2_-(x)} \,
     \frac{\delta B_a^{\bullet}(x^+;\mathbf{x})}{\delta {A}_c^{\bullet}(x^+;\mathbf{y})} \right] 
     {A}_c^{\star}(x^+;\mathbf{y})
      \, ,
    \label{eq:Bstar_Bbullet}
\end{equation}
where $\partial_-(x)=\partial/\partial x^-$.
From the above formula it is easy to see that the momentum space
expression (\ref{eq:Bstar}) with the kernels (\ref{eq:Upsilon_n}) 
indeed satisfies (\ref{eq:Bstar_WL}). The differential operator
$\partial^2_-(y)/\partial^2_-(x)$ will give the momentum dependent
prefactor in (\ref{eq:Upsilon_n}), whereas the functional derivative will replace one $A^{\bullet}$ by $A^{\star}$ in the Wilson line $B^{\bullet}$. For $n$-th term in the expansion there are $n$ possibilities to do so, thus the factor of $n$ in (\ref{eq:Upsilon_n}).

\begin{figure}
    \centering
    \includegraphics[width=9cm]{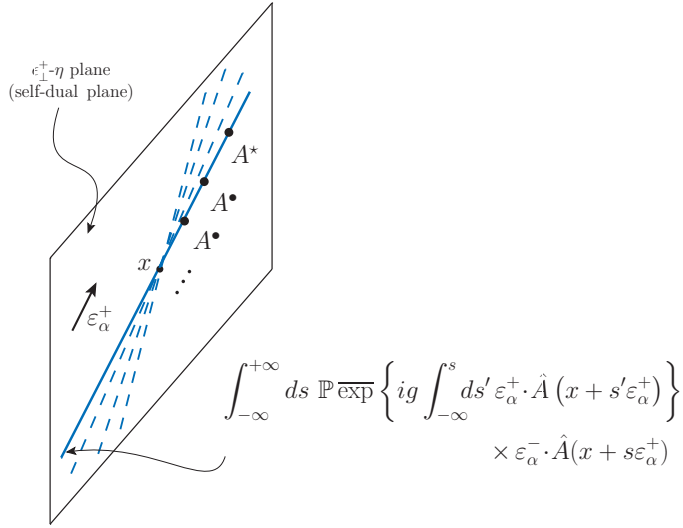}
    \caption{
    \small
    The solution to the $B^{\star}$ field can be represented as the straight infinite Wilson line similar to the one from Fig.~\ref{fig:SDplane}, but where one $A^{\bullet}$ field has been replaced by the $A^{\star}$ field. Alternatively, one can patch the semi-infinite Wilson line with the $A^{\star}$ field, as shown in the figure. The bar over $\exp$ function denotes the functional incorporating the $n$ factor in series expansion, i.e. $\overline{\exp}(x)=x\exp(x)$  
    } 
    \label{fig:SDplane_Bstar}
\end{figure}

Geometrically, the functional (\ref{eq:Bstar_WL}) is just the infinite Wilson line patched with one $A^{\star}$ field -- see Fig.~\ref{fig:SDplane_Bstar}. 
It is interesting to note, that it is natural to think about the $A^{\star}$ fields as belonging to Wilson lines living within the anti-self-dual plane spanned by $\varepsilon^-_{\alpha}$ 
(recall thet the $B^{\bullet}$ lives on the plane spanned by $\varepsilon^+_{\alpha}$).
Therefore, the solution (\ref{eq:Bstar_WL}) looks a bit like a cut through a bigger structure, spanning over both planes. Such possible extension of the Wilson line application in MHV formalism is beyond the scope of the present paper and is left for future study.

In the end of this subsection, let us note that from the equation (\ref{eq:Bstar_Bbullet}) one can derive another interesting formula for the $B^{\bullet}$ field.
Differentiating (\ref{eq:Bstar_Bbullet}) over $\partial_-(x)$ and using the Mansfield's transformation rule for the $A^{\star}$ field \cite{Mansfield2006}:
\begin{equation}
    \partial_-  A^{\star}_{a} (x) = 
    \int\! d^3\mathbf{y}\, \frac{\delta B^{\bullet}_c (x^+;\mathbf{y}) }{\delta A^{\bullet}_a (x^+;\mathbf{x}) }
    \, \partial_-  B^{\star}_{c} (x^+;\mathbf{y})
    \label{eq:Mansfield_Bstar}
\end{equation}
we easily obtain the following equation:
\begin{equation}
    \int\! d^3\mathbf{y} \,
     \left[ \partial^{-1}_-(x) \,
     \frac{\delta \hat{B}^{\bullet}(x^+;\mathbf{x})}{\delta \hat{A}^{\bullet}(x^+;\mathbf{y})}\right] \, 
     \left[ \partial_-(y)\,
     \frac{\delta \hat{B}^{\bullet}(x^+;\mathbf{z})}{\delta \hat{A}^{\bullet}(x^+;\mathbf{y})} \right]
     = \delta^3\left(\mathbf{x}-\mathbf{z}\right)
      \, .
    \label{eq:Bbullet_newEq}
\end{equation}
Integrating by parts over $y^-$ we see that merely the operator $\partial_-(y)/\partial_-(x)$ is responsible for inverting the functional derivative of the Wilson line.

\section{Summary}
\label{sec:Summary}

In this paper we have explored structures in the Yang-Mills theory that give rise to the MHV vertices in the CSW formulation. These structures turn out to be functionals of the gauge fields, directly related to the straight infinite Wilson lines extending over a complex self-dual plane spanned by  $\varepsilon_{\alpha}^+$ defined in Eq.~(\ref{eq:epsilon_alpha}).
Similar Wilson lines appear in the literature in the context of the gauge invariant  amplitudes with some external partons being kept off-shell. 

The functionals that involve complexified, straight, infinite Wilson lines transform the original fields appearing in the light cone Yang-Mills action to the fields which enter the MHV action. The light cone action involves just two components of the gauge field that correspond to plus and minus helicity gluons. The tranformation of the plus helicity field is given solely by the solution to the Self Dual equation of motion.
Therefore, we explored in details the connection between the Wilson line and the Self Dual sector of the Yang-Mills theory. To be precise we showed that the Wilson line expression satisfies the self-dual EOM, when the currents are restricted to the support on the light-cone. Thus the Wilson line itself represents a more general object than the self-dual current. We have also  found the explicit compact form of the transformation for the minus helicity field in position space. This turns out to be given by the functional derivative of the Wilson line solution, which effectively replaces one of the plus fields along the line by the minus field.

There are further  possible avenues to follow. One of them is to investigate further the geometry of the solutions. The Wilson lines corresponding to the positive helicity fields live on the plane spanned by the $\varepsilon_{\alpha}^+$. On the other hand one could think of the minus helicity fields as belonging to the gauge links which live in the plane spanned by $\varepsilon_{\alpha}^-$. Since the transformation for the minus helicity field that takes the Yang-Mills theory to the MHV action involves both plus and minus helicity fields, it appears to be a certain cut through a bigger geometrical object which is spanned over both planes. Investigation of such possible extensions of Yang-Mills field transformations is left for the future.



\section{Acknowledgments}
\label{sec:Acknowledgments}
HK and PK are supported by the National Science Center, Poland grant no. 2018/31/D/ST2/02731.  A.M.S. is supported  by the U.S. Department of Energy Grant 
 DE-SC-0002145 and  in part by  National Science Centre in Poland, grant 2019/33/B/ST2/02588.

\bibliographystyle{JHEP}
\bibliography{library}


\newpage
\appendix

\section{Vanishing of higher order terms in the Wilson line expansion in the on-shell limit}
\label{sec:AppA}

In this appendix, we prove that 
\begin{equation}
j_{a}[A]\left(x\right)=\int_{-\infty}^{\infty}d\alpha\,\mathrm{Tr}\left\{ \frac{1}{2\pi g}t^{a}\partial_{-}\, \mathbb{P}\exp\left[ig\int_{-\infty}^{\infty}ds\, \varepsilon_{\alpha}^+\cdot \hat{A}\left(x+s\varepsilon_{\alpha}^+\right)\right]\right\} \, ,
\label{eq:WilsonLineSol1}
\end{equation}
satisfies the self-dual EOM
\begin{equation}
    \Box {\hat{A}}^{\bullet} + 2ig{\partial}_{-} \left[ ({\partial}_{-}^{-1} {\partial}_{\bullet} {\hat{A}}^{\bullet}), {\hat{A}}^{\bullet} \right] - \Box\, \hat{j} = 0\, .
    \label{eq:SD_EOM_j1}
\end{equation}

To this end, we first Fourier transform (\ref{eq:SD_EOM_j1})\footnote{We follow the Fourier transform convention as in \cite{Kotko2017}} 
\begin{multline}
    -P^{2} \tilde{A}^{\bullet}_a (P) + ig f^{a b c} \int\!d^{4}p_{1}d^{4}p_{2} {\delta}^{4}(p_{1}+p_{2}-P) \left \{ \frac{p_{12}^{+}}{p_{1}^{+}} \times {\tilde v}_{12} \right \} \\
    \times \tilde{A}^{\bullet }_b (p_{1})\tilde{A}^{\bullet }_c (p_{2}) + P^{2} \tilde{j}_a (P) \, .
    \label{eq:eomft}
\end{multline}
where $P^{2} = 2(P^+ P^- - P^\bullet P^\star)$. Recall that the self-dual solution is expressed in terms of $j_a$ currents having the support \emph{on the light cone}. Thus we need to assume $P^2\rightarrow 0$. 

The Fourier transform of (\ref{eq:WilsonLineSol1}) is 
\begin{multline}
    \tilde{j}_a(P) = \sum_{n=1}^{\infty} 
    \int d^4p_1\dots d^4p_n \, \tilde{\Gamma}_n^{a\{b_1\dots b_n\}}(P;\{p_1,\dots ,p_n\}) \\ \tilde{A^{\bullet}}_{b_1}(p_1)\dots \tilde{A^{\bullet}}_{b_n}(p_n)\,.
    \label{eq:j[A]1}
\end{multline}
where 
\begin{equation}
    \tilde{\Gamma}_n^{a\{b_1\dots b_n\}} (P;\{p_1,\dots,p_n\}) = (-g)^{n-1} \frac{\delta^4\left(p_1+\dots+p_n-P\right)\, \mathrm{Tr} \left(t^{a} t^{b_{1}} \cdots t^{b_{n}}\right)}
    {\Tilde{v}^*_{1(1\cdots n)} \Tilde{v}^*_{(12)(1\cdots n)} \cdots \Tilde{v}^*_{(1 \cdots n-1)(1\cdots n)}} \, .
    \label{eq:Gamma_n1}
\end{equation}
Above is the 4D version of the Wilson line kernel (\ref{eq:Gamma_n}), \emph{cf.} (\ref{eq:Psi_n_3D}).
In what follows we skip the color indices for more compact formulae. 
For further use we can write
\begin{equation}
    P^2  \tilde{\Gamma}_n(p_1 , p_2 , \cdots p_n)    = 
     -P^+ D_{1 \cdots n} \tilde{\Gamma}_n(p_1 , p_2 , \cdots p_n) \, ,
    \label{eq:pd}
\end{equation}
where
\begin{equation}
    D_{1 \cdots n} =  2(\frac{P^\bullet P^\star}{P^+} - P^-)=
    \frac{1}{P^+} \sum_{i,j=1}^{n} \Tilde{v}_{ij}\Tilde{v}^*_{ji} \, ,
    \label{eq:vvstar}
\end{equation}
where we have used an identity similar to the Schouten identity for spinor products in the on-shell case. The above expression holds because the momenta $p_1,\dots,p_n$ are on-shell.

Let us now proceed term by term in (\ref{eq:eomft}).  
Since the first term in (\ref{eq:j[A]1}) is $\tilde{A}^{\bullet}_a(P)$, it cancels out the first term of (\ref{eq:eomft}). For the second term in (\ref{eq:j[A]1}), we have
\begin{equation}
   \tilde{\Gamma}_2(p_1 , p_2) = -g \frac{{\delta}^{4}(p_{1}+p_{2}-P)}{{\tilde v}^{\ast}_{1(12)}} \mathrm{Tr}(t^{a} t^{b_{1}} t^{b_{2}}) \, .
\end{equation}
It is easy to show that the above expression can be written as
\begin{equation}
   \tilde{\Gamma}_2(p_1 , p_2) =   \frac{{\delta}^{4}(p_{1}+p_{2}-P)}{{\tilde v}_{21}{\tilde v}^{\ast}_{12}} \frac{ig}{2}f^{c b_{1}b_{2}}\left \{ \frac{p_{12}^{+}}{p_{1}^{+}} \times {\tilde v}_{12} \right \} \, .
\end{equation}
Using (\ref{eq:vvstar}) we get
\begin{equation}
     P^2  \tilde{\Gamma}_2(p_1 , p_2) = -igf^{c b_{1}b_{2}}\left \{ \frac{p_{12}^{+}}{p_{1}^{+}} \, {\tilde v}_{12} \right \}{\delta}^{4}(p_{1}+p_{2}-P) \, .
     \label{eq:2gam}
\end{equation}
Comparing this with the second term of (\ref{eq:eomft}), we see that they cancel out. An important point to note is that equation (\ref{eq:2gam}) represents a 3-gluon vertex where the incoming gluon is off-shell and the outgoing gluons are on-shell. When we impose the on-shell condition, $P^2 \rightarrow 0$, the quantity on the right side of (\ref{eq:2gam}) represents 3-gluon vertex with all gluons on-shell. This we know must be zero. Hence we have,
\begin{equation}
    {\tilde v}_{12} \rightarrow 0 \, .
    \label{eq:3gv0}
\end{equation}

Since we have canceled the two terms present in (\ref{eq:SD_EOM_j1}) (the $g^0$ and $g^1$ terms), we now need to show that
\begin{equation}
     \sum_{n=3}^{\infty}\left [\int\!d^{4}p_{1}\cdots d^{4}p_{n} P^2 \tilde{\Gamma}_{n}(p_1 , p_2 , \cdots p_n) {\tilde A}^{\bullet} (p_1) {\tilde A}^{\bullet} (p_2)\cdots  {\tilde A}^{\bullet} (p_n) \right ]_{P^2 \rightarrow 0} = 0 \, .
     \label{eq:hotz}
\end{equation}
 From equation  (\ref{eq:vvstar}) we have
\begin{eqnarray}
    P^2 = -\sum_{i,j=1}^{n} \Tilde{v}_{ij}\Tilde{v}^*_{ji} = \sum_{i,j=1}^{n} \frac{p_j^+}{p_i^+}\Tilde{v}_{ij}\Tilde{v}^*_{ij} \, .
    \label{eq:pode}
\end{eqnarray}
Above, we see that each term in the expansion of $P^2$ is positive definite. Hence under the limit $P^2 \rightarrow 0$, each term in  (\ref{eq:pode}) must independently go to zero. Thus
\begin{equation}
  P^2 \rightarrow 0 \hspace{0.5cm}\Rightarrow \hspace{0.5cm} \Tilde{v}_{ij}\Tilde{v}^*_{ji} \longrightarrow 0 \hspace{0.5cm} \forall i,j \, .
  \label{eq:vvstarz}
\end{equation}
The implication of the expression above (which will be useful later) is the following.
Since the three gluon vertex for splitting $P\rightarrow p_i p_j$, with $i,j$ on-shell, is proportional to $\tilde{v}_{ij}$ (\emph{cf.} \ref{eq:2gam}), we must conclude that the on-shell limit is approached by
\begin{equation}
    \tilde{v}_{ij}\rightarrow 0\, \hspace{0.5cm} \forall i,j.
    \label{eq:vtildeto0}
\end{equation}
This follows from the fact that for $n=2$ Eq.~(\ref{eq:pode}) is just $\sim\Tilde{v}_{12}\Tilde{v}^*_{21}$. On the other hand the fully on-shell 3-gluon vertex is $\sim \tilde{v}_{12}$ and must be zero due to momentum conservation. Thus, in the on-shell limit $\tilde{v}_{12}\rightarrow 0$. This holds for any 3-gluon vertex with on-shell legs.

In \cite{Kotko2017} it was shown that  $\tilde{\Gamma}_{n}(p_1 , p_2 , \cdots p_n)$ can be written as
\begin{multline}
     \tilde{\Gamma}_{n}(p_1 , p_2 , \cdots p_n) = (-g)^{n-1} \delta^{4} (p_{1} + \cdots +p_{n} - P) \frac{-2}{D_{1 \cdots n}} \\
    \times\!\frac{v_{n(n-1)}\Tilde{v}^*_{(1 \cdots n-1)(1\cdots n)} + v_{(n-1)(n-2)} \Tilde{v}^*_{(1 \cdots n-2)(1\cdots n)} + \cdots v_{21}\Tilde{v}^*_{1(1\cdots n)} }{\Tilde{v}^*_{1(1\cdots n)} \Tilde{v}^*_{(12)(1\cdots n)} \cdots \Tilde{v}^*_{(1 \cdots n-1)(1\cdots n)}} \, .
    \label{eq:gfac}
\end{multline}
where
\begin{equation}
    {v}_{ij}=\frac{1}{p^+_j}\tilde{v}_{ji}=
    \left(\frac{p_{i}^{\star}}{p_{i}^{+}}-\frac{p_{j}^{\star}}{p_{j}^{+}}\right) \, .
\label{eq:vij}
\end{equation}
Using this, we can write
\begin{multline}
     P^2\tilde{\Gamma}_{n}(p_1 , p_2 , \cdots p_n) = (-g)^{n-1}\delta^{4} (p_{1} + \cdots +p_{n} - P) 2P^+ \\
    \times\!\frac{v_{n(n-1)}\Tilde{v}^*_{(1 \cdots n-1)(1\cdots n)} + v_{(n-1)(n-2)} \Tilde{v}^*_{(1 \cdots n-2)(1\cdots n)} + \cdots v_{21}\Tilde{v}^*_{1(1\cdots n)} }{\Tilde{v}^*_{1(1\cdots n)} \Tilde{v}^*_{(12)(1\cdots n)} \cdots \Tilde{v}^*_{(1 \cdots n-1)(1\cdots n)}} \, .
\end{multline}

It can be rewritten as 
\begin{multline}
     P^2\tilde{\Gamma}_{n}(p_1 , p_2 , \cdots p_n) = (-g)^{n-1}\delta^{4} (p_{1} + \cdots +p_{n} - P) 2P^+  \\
    \times \left [ \frac{v_{n(n-1)}}{\Tilde{v}^*_{1(1\cdots n)} \Tilde{v}^*_{(12)(1\cdots n)} \cdots \Tilde{v}^*_{(1 \cdots n-2)(1\cdots n)}} \right. \\
    \left. + \frac{v_{(n-1)(n-2)}}{\Tilde{v}^*_{1(1\cdots n)} \Tilde{v}^*_{(12)(1\cdots n)} \cdots \Tilde{v}^*_{(1 \cdots n-3)(1\cdots n)}\Tilde{v}^*_{(1 \cdots n-1)(1\cdots n)}} + \cdots \right] \, .
\end{multline}
Using Eqs.~(\ref{eq:3gv0}),(\ref{eq:vij}) we see that each term in the above expression is zero under the limit $P^2 \rightarrow 0$. 
Therefore the whole expression vanishes in that limit.

\section{Prove of the $B^{\star}$ field expansion in momentum space}
\label{sec:AppB}

In this appendix we derive the momentum space power series for ${\tilde{B}}^\star (x^+;\mathbf{P})$ which we postulate to have the following expansion
\begin{multline}
    {\tilde{B}}_a^\star (x^+;\mathbf{P}) = {\tilde A}^\star_{a} (x^+;\mathbf{P}) + \sum_{n=2}^{\infty} \int\!d^{3}\mathbf{p_1}\cdots d^{3}\mathbf{p_n} {\tilde \Upsilon}_{n}^{a b_1 \left \{b_2 \cdots b_n \right \} }(\mathbf{P}; \mathbf{p_1} ,\left \{ \mathbf{p_2} , \cdots \mathbf{p_n} \right \})  \\
     {\tilde A}^\star_{b_1} (x^+;\mathbf{p_1}){\tilde A}^\bullet_{b_2} (x^+;\mathbf{p_2}) \cdots {\tilde A}^\bullet_{b_n} (x^+;\mathbf{p_n}) \, ,
   \label{eq:Bstar1}
\end{multline}
where, throughout the derivation, the curly braces represent that the enclosed momenta and color indices are symmetrized. The goal is to find the coefficient functions ${\tilde \Upsilon}_{n}$ which we shall derive for a first few terms and generalize the result. 

The ${\tilde A}^\bullet$ and ${\tilde A}^\star$ fields were shown in \cite{Kotko2017} to have the following power series expansions:
\begin{multline}
     {\tilde {A}}^{\bullet}_{a} (x^+;\mathbf{P}) = {\tilde B}^{\bullet}_{a} (x^+;\mathbf{P}) + \sum_{n=2}^{\infty} \int\!d^{3}\mathbf{p_{1}}\cdots d^{3}\mathbf{p_{n}} {\tilde \Psi}_{n}^{a \left \{b_1 \cdots b_n \right \} }(\mathbf{P}; \left \{\mathbf{p_{1}}  \cdots \mathbf{p_{n}} \right \}) \\  
     {\tilde B}^{\bullet}_{b_1} (x^+;\mathbf{p_{1}}){\tilde B}^{\bullet}_{b_2} (x^+;\mathbf{p_{2}}) \cdots {\tilde B}^{\bullet}_{b_n}(x^+;\mathbf{p_{n}}) \, ,
   \label{eq:abullet}
\end{multline}
where
\begin{multline}
    {\tilde \Psi}_{n}^{a \left \{b_1 \cdots b_n \right \} }(\mathbf{P}; \left \{\mathbf{p_{1}}  \cdots \mathbf{p_{n}} \right \}) =- (-g)^{n-1}  \frac{{\tilde v}^{\ast}_{(1 \cdots n)(1)}}{{\tilde v}^{\ast}_{(2)(1)}{\tilde v}^{\ast}_{(3)(2)} \cdots {\tilde v}^{\ast}_{(n)(n-1)}{\tilde v}^{\ast}_{(1)(1 \cdots n)}}  \\
    \times \delta^{3} (\mathbf{p_{1}} + \cdots +\mathbf{p_{n}} - \mathbf{P}) \mathrm{Tr} (t^{a} t^{b_{1}} \cdots t^{b_{n}}) \, ,
    \label{eq:wspsi}
\end{multline}
and
\begin{multline}
    {\tilde A}^{\star}_{a} (x^+;\mathbf{P}) = {\tilde B}^{\star}_{a} (x^+;\mathbf{P}) +
    \sum_{n=2}^{\infty} \int\!d^{3}\mathbf{p_{1}}\cdots d^{3}\mathbf{p_{n}} {\tilde \Omega}_{n}^{a b_1 \left \{b_2 \cdots b_n \right \} }(\mathbf{P}; \mathbf{p_{1}} ,\left \{ \mathbf{p_{2}} , \cdots \mathbf{p_{n}} \right \}) \\ {\tilde B}^{\star}_{b_1} (x^+;\mathbf{p_{1}}) {\tilde B}^{\bullet}_{b_2} (x^+;\mathbf{p_{2}})\cdots {\tilde B}^{\bullet}_{b_n}(x^+;\mathbf{p_{n}}) \, ,
    \label{eq:astar}
\end{multline}
with
\begin{equation}
    {\tilde \Omega}_{n}^{a b_1 \left \{b_2 \cdots b_n \right \} }(\mathbf{P}; \mathbf{p_{1}} , \left \{ \mathbf{p_{2}} , \cdots \mathbf{p_{n}} \right \} ) = n \left(\frac{p_1^+}{p_{1\cdots n}^+}\right)^2 {\tilde \Psi}_{n}^{a b_1 \cdots b_n }(\mathbf{P};  \mathbf{p_{1}}  \cdots \mathbf{p_{n}} ) \, .
    \label{eq:pocor}
\end{equation}
The kernel ${\tilde \Psi}_{n}^{a  b_1 \cdots b_n }(\mathbf{P}; \mathbf{p_{1}}  \cdots \mathbf{p_{n}} )$ can be  written as
\begin{equation}
    {\tilde \Psi}_{n}^{a b_1 \cdots b_n }(\mathbf{P}; \mathbf{p_{1}}  \cdots \mathbf{p_{n}} ) = \sum_{\mathrm{permutations}} \mathrm{Tr} (a 1 \cdots n)\Psi_n (1 \cdots n) \, .
    \label{eq:psiperm}
\end{equation}
where the sum is over all permutations of ($1 \cdots n$). Above, we have introduced the following notation for convenience 
\begin{eqnarray}
    \mathrm{Tr} (a 1 \cdots n) &=& \mathrm{Tr} (t^{a} t^{b_{1}} \cdots t^{b_{n}}) \, ,  \label{eq:psiperm1} \\
    \Psi_n (1 \cdots n) &=& - \frac{1}{n!}(-g)^{n-1}  \frac{{\tilde v}^{\ast}_{(1 \cdots n)(1)}}{{\tilde v}^{\ast}_{(2)(1)}{\tilde v}^{\ast}_{(3)(2)} \cdots {\tilde v}^{\ast}_{(n)(n-1)}{\tilde v}^{\ast}_{(1)(1 \cdots n)}} \nonumber\\
    && \times \delta^{3} (\mathbf{p_{1}} + \cdots +\mathbf{p_{n}} - \mathbf{P}) \, , \label{eq:psiperm2}
\end{eqnarray}
The same notations will be used for the coefficient functions of the power series for ${\tilde B}^\bullet$ in the momentum space, ${\tilde \Gamma}_{n}^{a b_1 \cdots b_n  }(\mathbf{P}; \mathbf{p_{1}}  \cdots \mathbf{p_{n}} )$.

To obtain  ${\tilde \Upsilon}_{n}$ we substitute equations (\ref{eq:astar}) and (\ref{eq:abullet}) in (\ref{eq:Bstar1}) and equate the terms of equal order. To differentiate the color indices and momentum variables, we use $c_1 , c_2, \cdots c_n$ as the color indices and $q_1 , q_2, \cdots q_n$ as momentum for (\ref{eq:Bstar1}) and respectively  $d_1 , d_2, \cdots d_n$ and $s_1 , s_2, \cdots s_n$ for (\ref{eq:abullet}) and  $b_1 , b_2, \cdots b_n$ and $p_1 , p_2, \cdots p_n$ for (\ref{eq:astar}).

The first order term is trivial. For the second order, we get
\begin{multline}
    0 = \int\!d^{3}\mathbf{q_{1}}d^{3}\mathbf{q_{2}} {\tilde \Upsilon}_{2}^{a c_1 \left \{c_2 \right \} }(\mathbf{P}; \mathbf{q_{1}} ,\left \{ \mathbf{q_{2}} \right \}) {\tilde B}^\star_{c_1} (x^+;\mathbf{q_{1}}){\tilde B}^\bullet_{c_2} (x^+;\mathbf{q_{2}})  \\
 +  \int\!d^{3}\mathbf{p_{1}}d^{3}\mathbf{p_{2}} {\tilde \Omega}_{2}^{a b_1 \left \{b_2 \right \} }(\mathbf{P}; \mathbf{p_{1}} ,\left \{ \mathbf{p_{2}} \right \}) {\tilde B}^{\star}_{b_1} (x^+;\mathbf{p_{1}}) {\tilde B}^{\bullet}_{b_2} (x^+;\mathbf{p_{2}})\, .
  \label{eq:2ndterm}
\end{multline}
Substituting for ${\tilde \Omega}_{2}$ using expression (\ref{eq:pocor}) for $n=2$ we have
\begin{equation}
    {\tilde \Upsilon}_{2}^{a b_1 \left \{b_2 \right \} }(\mathbf{P}; \mathbf{p_{1}} ,\left \{ \mathbf{p_{2}} \right \})  = - 2 \left(\frac{p_1^+}{p_{12}^+}\right )^2 \left[ \Psi_2(12)\mathrm{Tr}(a12) + \Psi_2(21)\mathrm{Tr}(a21)\right] \, .
    \label{eq:uppsi}
\end{equation}
Using the result $\Psi_2 = -\Gamma_2$ from \cite{Kotko2017} we obtain
\begin{equation}
    {\tilde \Upsilon}_{2}^{a b_1 \left \{b_2 \right \} }(\mathbf{P}; \mathbf{p_{1}} ,\left \{ \mathbf{p_{2}} \right \})  =  2 \left(\frac{p_1^+}{p_{12}^+}\right )^2  {\tilde \Gamma}_{2}^{a b_1 b_2 }(\mathbf{P}; \mathbf{p_{1}} , \mathbf{p_{2}} ) \, .
    \label{eq:r2ob}
\end{equation}

For the third order term, we proceed analogously. We have
\begin{multline}
     0 = \int\!d^{3}\mathbf{q_{1}}d^{3}\mathbf{q_{2}} d^{3}\mathbf{q_{3}} {\tilde \Upsilon}_{3}^{a c_1 \left \{c_2 c_3 \right \} }(\mathbf{P}; \mathbf{q_{1}} ,\left \{ \mathbf{q_{2}} , \mathbf{q_{3}} \right \}) {\tilde B}^\star_{c_1} (x^+;\mathbf{q_{1}}){\tilde B}^\bullet_{c_2} (x^+;\mathbf{q_{2}})\\
{\tilde B}^\bullet_{c_3} (x^+;\mathbf{q_{3}})  + \int\!d^{3}\mathbf{p_{1}}d^{3}\mathbf{p_{2}} d^{3}\mathbf{p_{3}} {\tilde \Omega}_{3}^{a b_1 \left \{b_2 b_3 \right \} }(\mathbf{P}; \mathbf{p_{1}} ,\left \{ \mathbf{p_{2}} , \mathbf{p_{3}} \right \})  {\tilde B}^{\star}_{b_1} (x^+;\mathbf{p_{1}})  \\
 {\tilde B}^{\bullet}_{b_2} (x^+;\mathbf{p_{2}}) {\tilde B}^{\bullet}_{b_3}(x^+;\mathbf{p_{3}})+ \int\!d^{3}\mathbf{q_{1}}d^{3}\mathbf{q_{2}} {\tilde \Upsilon}_{2}^{a c_1 \left \{c_2 \right \} }(\mathbf{P} ; \mathbf{q_{1}} ,\left \{ \mathbf{q_{2}} \right \})  \\
 \left[\int\!d^{3}\mathbf{p_{1}}d^{3}\mathbf{p_{2}} {\tilde \Omega}_{2}^{c_1 b_1 \left \{b_2 \right \} }(\mathbf{q_1}; \mathbf{p_{1}},\left \{ \mathbf{p_{2}}  \right \}) {\tilde B}^\star_{b_1} (x^+;\mathbf{p_{1}}){\tilde B}^\bullet_{b_2} (x^+;\mathbf{p_{2}}){\tilde B}^{\bullet}_{c_2} (x^+;\mathbf{q_{2}}) \right.\\
 \left. + {\tilde B}^{\star}_{c_1} (x^+;\mathbf{q_{1}})\int\!d^{3}\mathbf{s_{1}} d^{3}\mathbf{s_{2}} {\tilde \Psi}_{2}^{c_2  \left \{d_1 d_2 \right \} }(\mathbf{q_{2}}; \left \{ \mathbf{s_{1}} , \mathbf{s_{2}} \right \}) {\tilde B}^\bullet_{d_1} (x^+;\mathbf{s_{1}}){\tilde B}^\bullet_{d_2} (x^+;\mathbf{s_{2}})    \right]
\end{multline}
Integrating the third term with respect to $\mathbf{q_{1}}$ and the fourth term with respect to $\mathbf{q_{2}}$  we see each term will have only three momentum variables. All the terms can be combined into one integral by renaming the momentum variables to  $\mathbf{t_{1}},\mathbf{t_{2}},\mathbf{t_{3}}$, and color indices to $e_1$, $e_2$, $e_3$. With this we have
\begin{multline}
    0 = \int\!d^{3}\mathbf{t_{1}}d^{3}\mathbf{t_{2}} d^{3}\mathbf{t_{3}}  {\tilde B}^\star_{e_1} (x^+;\mathbf{t_{1}}){\tilde B}^\bullet_{e_2} (x^+;\mathbf{t_{2}}){\tilde B}^\bullet_{e_3} (x^+;\mathbf{t_{3}})  \\
 \left[ {\tilde \Upsilon}_{3}^{a e_1 \left \{e_2 e_3 \right \} }(\mathbf{P}; \mathbf{t_{1}} ,\left \{ \mathbf{t_{2}} , \mathbf{t_{3}} \right \})  + {\tilde \Omega}_{3}^{a e_1 \left \{e_2 e_3 \right \} }(\mathbf{P}; \mathbf{t_{1}} ,\left \{ \mathbf{t_{2}} , \mathbf{t_{3}} \right \})  \right. \\
 \left. + {\tilde \Upsilon}_{2}^{a c_1 \left \{e_3 \right \} }(\mathbf{P} ; \mathbf{t_{1}}+\left\{ \mathbf{t_{2}}\right\} ,\left \{ \mathbf{t_{3}} \right \}){\tilde \Omega}_{2}^{c_1 e_1 \left \{e_2 \right \} }(\mathbf{q_1}; \mathbf{t_{1}},\left \{ \mathbf{t_{2}}  \right \})\right. \\
 \left.  + {\tilde \Upsilon}_{2}^{a e_1 \left \{c_2 \right \} }(\mathbf{P} ; \mathbf{t_{1}} ,\left \{ \mathbf{t_{2}}+\mathbf{t_{3}} \right \}) {\tilde \Psi}_{2}^{c_2  \left \{e_2 e_3 \right \} }(\mathbf{q_{2}}; \left \{ \mathbf{t_{2}} , \mathbf{t_{3}} \right \})
 \right]
\end{multline}
Substituting for ${\tilde \Omega}_{3}$, ${\tilde \Omega}_{2}$ and ${\tilde \Upsilon}_{2}$ we get
\begin{multline}
    {\tilde \Upsilon}_{3}^{a e_1 \left \{e_2 e_3 \right \} }(\mathbf{P}; \mathbf{t_{1}} ,\left \{ \mathbf{t_{2}} , \mathbf{t_{3}} \right \}) = - \left[{\tilde \Omega}_{3}^{a e_1 \left \{e_2 e_3 \right \} }(\mathbf{P}; \mathbf{t_{1}} ,\left \{ \mathbf{t_{2}} , \mathbf{t_{3}} \right \})  \right.  \\
 \left.  + {\tilde \Upsilon}_{2}^{a c_1 \left \{e_3 \right \} }(\mathbf{P} ; \mathbf{t_{1}}+\left\{ \mathbf{t_{2}}\right\} ,\left \{ \mathbf{t_{3}} \right \}) {\tilde \Omega}_{2}^{c_1 e_1 \left \{e_2 \right \} }(\mathbf{q_1}; \mathbf{t_{1}},\left \{ \mathbf{t_{2}}  \right \})  \right.  \\
 \left. + {\tilde \Upsilon}_{2}^{a e_1 \left \{c_2 \right \} }(\mathbf{P} ; \mathbf{t_{1}} ,\left \{ \mathbf{t_{2}}+\mathbf{t_{3}} \right \}){\tilde \Psi}_{2}^{c_2  \left \{e_2 e_3 \right \} }(\mathbf{q_{2}}; \left \{ \mathbf{t_{2}} , \mathbf{t_{3}} \right \})
 \right] \,.
 \label{eq:up3lon}
\end{multline}

In order to proceed, we introduce the following notation  
\begin{equation}
    \Psi_n^B (1 \cdots n) = (n!) \, \Psi_n (1 \cdots n) \, 
    \label{eq:nottr}
\end{equation}
and similar for $\tilde{\Gamma}_n$. This will be necessary to decouple the symmetry factors $n!$ hidden in the kernels from the dynamical part.
 Using the notation (\ref{eq:nottr}), expression (\ref{eq:up3lon}) can be written, after a bit of algebra, as
 \begin{multline}
     {\tilde \Upsilon}_{3}^{a e_1 \left \{e_2 e_3 \right \} }(\mathbf{P}; \mathbf{t_{1}} ,\left \{ \mathbf{t_{2}} , \mathbf{t_{3}} \right \})
 =  \left(\frac{t_1^+}{t^+_{123}}\right)^2\frac{1}{2}\\ 
\times\left[ \left(-\Psi_3^B(123)+ \Gamma_2^B(\left[12\right]3) \Gamma_2^B(12) +\Gamma_2^B(1\left[23\right]) \Gamma_2^B(23) \right)\mathrm{Tr}(a123)  \right.  \\
 \left. +\left( -\Psi_3^B(132)+ \Gamma_2^B(\left[13\right]2) \Gamma_2^B(13) + \Gamma_2^B(1\left[32\right]) \Gamma_2^B(32) \right)\mathrm{Tr}(a132) \right. \\
\left. +\left( -\Psi_3^B(213)+ \Gamma_2^B(\left[21\right]3) \Gamma_2^B(21) + \Gamma_2^B(2\left[13\right]) \Gamma_2^B(13) \right)\mathrm{Tr}(a213) \right. \\
 \left. +\left(-\Psi_3^B(231)+ \Gamma_2^B(\left[23\right]1) \Gamma_2^B(23) + \Gamma_2^B(2\left[31\right]) \Gamma_2^B(31) \right)\mathrm{Tr}(a231)\right. \\
\left.+\left( -\Psi_3^B(312)+\Gamma_2^B(\left[31\right]2) \Gamma_2^B(31) + \Gamma_2^B(3\left[12\right]) \Gamma_2^B(12) \right)\mathrm{Tr}(a312) \right.\\
\left. +\left(-\Psi_3^B(321)+ \Gamma_2^B(\left[32\right]1) \Gamma_2^B(32) + \Gamma_2^B(3\left[21\right]) \Gamma_2^B(21) \right)\mathrm{Tr}(a321) \right] \, .
\label{eq:up3gam3}
 \end{multline}
where $\left[ij\right] = \mathbf{t_{i}} + \mathbf{t_{j}}$.
From \cite{Kotko2017} we note that each term on the right hand side is a permutation of $\Gamma_3^B$.

The expression (\ref{eq:up3gam3}) can thus be compactly written as
\begin{equation}
    {\tilde \Upsilon}_{3}^{a e_1 \left \{e_2 e_3 \right \} }(\mathbf{P}; \mathbf{t_{1}} ,\left \{ \mathbf{t_{2}} , \mathbf{t_{3}} \right \})
 =  3\left(\frac{t_1^+}{t^+_{123}}\right)^2  {\tilde \Gamma}_3^{a e_1 e_2 e_3}\left(\mathbf{P}; \mathbf{t_{1}} , \mathbf{t_{2}} , \mathbf{t_{3}} \right ) \, .
\end{equation}

This can be readily generalized for any $n$ as
\begin{equation}
    {\tilde \Upsilon}_{n}^{a b_1 \left \{b_2 \cdots b_n \right \} }(\mathbf{P}; \mathbf{p_1} ,\left \{ \mathbf{p_2} , \cdots \mathbf{p_n} \right \}) = n\left(\frac{p_1^+}{p_{1\cdots n}^+}\right )^2 {\tilde \Gamma}_{n}^{a b_1 \cdots b_n }\left(\mathbf{P}; \mathbf{p_{1}}  \cdots \mathbf{p_{n}} \right) 
\end{equation}

\section{Prove of the expression for $B^{\star}$ in terms of the Wilson line derivative}
\label{sec:AppC}

In this appendix we prove Eq.~(\ref{eq:Bstar_WL}), which we repeat below for convenience
\begin{multline}
    B_a^{\star}(x) = 
    \int_{-\infty}^{+\infty}\! d\alpha\,\, 
    \mathrm{Tr} \Big\{
    \frac{1}{2\pi g} t^a \partial_-^{-1} 
    \int\! d^4y \,
     \left[\partial_-^2 {A}^{\star}_c (y)\right]  \\
     \frac{\delta}{\delta {A}^{\bullet}_c (y)} \,
    \mathbb{P} \exp {\left[ig \int_{-\infty}^{+ \infty}\! ds \,
    \hat{A}^{\bullet}(x+s\varepsilon_{\alpha})\right]  } 
    \Big\} \,.
    \label{eq:Bstar_WL1}
\end{multline}
For convenience, let us rewrite (\ref{eq:Bstar_WL1}) as follows
\begin{equation}
    B_a^{\star}(x) = 
    \int_{-\infty}^{+\infty}\! d\alpha\,\, 
    \mathrm{Tr} \Big\{
    \frac{1}{2\pi g} t^a \partial_-^{-1} G(x)\Big\} \,.
    \label{eq:bstar2}
\end{equation}
where
\begin{equation}
    G(x) = \int\! d^4y \,
     \left[\partial_-^2 {A}^{\star}_c (y)\right]  
     \frac{\delta}{\delta {A}^{\bullet}_c (y)} \,
    \mathbb{P} \exp {\left[ig \int_{-\infty}^{+ \infty}\! ds \,
    \hat{A}^{\bullet}(x+s\varepsilon_{\alpha})\right]  } \, .
    \label{eq:gdef}
\end{equation}
The functional derivative of the path ordered integral will give a series containing in each term 
a Dirac delta ($\delta^4 (x+s_i \varepsilon_{\alpha}- y)$) for the position space variables and a Kronecker delta ($\delta^c_{b_i}$) for the color indices. Integrating with respect to $y$, we get
\begin{multline}
    G(x)= ig\int_{-\infty}^{+ \infty}\! ds_1\,  \partial_-^2 {\hat{A}}^\star (x+s_1 \varepsilon_{\alpha}) +  (ig)^2\int_{-\infty}^{+ \infty} \! ds_1\,\int_{-\infty}^{s_1}\!  ds_2\, \left \{{\hat{A}}^\bullet (x+s_1 \varepsilon_{\alpha})\right.   \\
    \left. \partial_-^2 {\hat{A}}^\star(x+s_2 \varepsilon_{\alpha}) + \partial_-^2 {\hat{A}}^\star(x+s_1 \varepsilon_{\alpha}){\hat{A}}^\bullet (x+s_2 \varepsilon_{\alpha}) \right \} + (ig)^3\int_{-\infty}^{+ \infty} \! ds_1\, \\
    \int_{-\infty}^{s_1} \! ds_2\, \int_{-\infty}^{s_2} \! ds_3\,  \left \{ {\hat{A}}^\bullet (x+s_1 \varepsilon_{\alpha}){\hat{A}}^\bullet (x+s_2 \varepsilon_{\alpha}) \partial_-^2 {\hat{A}}^\star(x+s_3 \varepsilon_{\alpha}) \right.   \\
    \left. + {\hat{A}}^\bullet (x+s_1 \varepsilon_{\alpha})\partial_-^2 {\hat{A}}^\star(x+s_2 \varepsilon_{\alpha}){\hat{A}}^\bullet (x+s_3 \varepsilon_{\alpha})  \right.   \\
    \left. +\partial_-^2 {\hat{A}}^\star(x+s_1 \varepsilon_{\alpha}) {\hat{A}}^\bullet (x+s_2 \varepsilon_{\alpha}){\hat{A}}^\bullet (x+s_3 \varepsilon_{\alpha})\right\} + \cdots 
\end{multline}
The above expression can be rewritten as
\begin{multline}
    G(x^+;\mathbf{x})= ig\int_{-\infty}^{+\infty}\! ds_1\,\int_{-\infty}^{+\infty}\!d^3\, \mathbf{p_1}(-i p_1^+)^2 e^{-i s_1 \mathbf{e}_{\alpha}\cdot \mathbf{p_1}} {\tilde A}_{b_1}^\star (x^+;\mathbf{p_1}) t^{b_1}  e^{-i\mathbf{x}\cdot\mathbf{p_1}} \\
+(ig)^2 \int_{-\infty}^{+\infty}\! ds_1\, \int_{-\infty}^{s_1} \! ds_2\, \int_{-\infty}^{+\infty}\! d^3\,\mathbf{p_1}\, d^3 \mathbf{p_2}\,e^{-i s_1 \mathbf{e}_{\alpha}\cdot \mathbf{p_1}}e^{-i s_2 \mathbf{e}_{\alpha}\cdot \mathbf{p_2}} e^{-i\mathbf{x}\cdot(\mathbf{p_1}+\mathbf{p_2})}  \\
 \times    \left\{ {\tilde A}_{b_1}^\bullet (x^+;\mathbf{p_1})t^{b_1}(-i p_2^+)^2{\tilde A}_{b_2}^\star (x^+;\mathbf{p_2}) t^{b_2} +    (-i p_1^+)^2{\tilde A}_{b_1}^\star (x^+;\mathbf{p_1})t^{b_1}\right. \\
\left. {\tilde A}_{b_2}^\bullet (x^+;\mathbf{p_2})t^{b_2}\right\} + (ig)^3\int_{-\infty}^{+ \infty}\! ds_1\,\int_{-\infty}^{s_1} \! ds_2\,\int_{-\infty}^{s_2} \! ds_3\,\int_{-\infty}^{+\infty}\!d^3\,\mathbf{p_1} d^3 \mathbf{p_2}\, d^3 \mathbf{p_3}\, \\
 \times e^{-i s_1 \mathbf{e}_{\alpha}\cdot \mathbf{p_1}}e^{-i s_2 \mathbf{e}_{\alpha}\cdot \mathbf{p_2}}  e^{-i s_3 \mathbf{e}_{\alpha}\cdot \mathbf{p_3}}e^{-i\mathbf{x}\cdot(\mathbf{p_1}+\mathbf{p_2}+\mathbf{p_3})} \\
 \times\left\{ {\tilde A}_{b_1}^\bullet (x^+;\mathbf{p_1})t^{b_1} {\tilde A}_{b_2}^\bullet (x^+;\mathbf{p_2})t^{b_2}(-i p_3^+)^2{\tilde A}_{b_3}^\star (x^+;\mathbf{p_3})t^{b_3}\right. \\
\left.  +  {\tilde A}_{b_1}^\bullet (x^+;\mathbf{p_1})t^{b_1}  (-i p_2^+)^2{\tilde A}_{b_2}^\star (x^+;\mathbf{p_2})t^{b_2}{\tilde A}_{b_3}^\bullet (x^+;\mathbf{p_3})t^{b_3}\right.  \\
\left. + (-i p_1^+)^2{\tilde A}_{b_1}^\star (x^+;\mathbf{p_1})t^{b_1} {\tilde A}_{b_2}^\bullet (x^+;\mathbf{p_2})t^{b_2}{\tilde A}_{b_3}^\bullet (x^+;\mathbf{p_3})t^{b_3}\right\} + \cdots
\label{eq:gdcop}
\end{multline}
where $\mathbf{e}_{\alpha} \equiv (-\alpha, 0, -1)$ are the  $(x^-, x^\bullet, x^\star)$ coordinates of  $\varepsilon_{\alpha}$. For the ordered integrals we have
\begin{multline}
     \int_{-\infty}^{+\infty}\! ds_1\, \cdots \int_{-\infty}^{s_{n-1}} \!ds_n \, e^{-i s_1 \mathbf{e}_{\alpha}\cdot \mathbf{p_1}}\cdots e^{-i s_n \mathbf{e}_{\alpha}\cdot \mathbf{p_n}} = 2\pi\delta(\mathbf{e}_{\alpha}\cdot \mathbf{p_{1\cdots n}}) \\
    \times \frac{i^{n-1}}{(\mathbf{e}_{\alpha}\cdot \mathbf{p_{2\cdots n}} + i\epsilon)(\mathbf{e}_{\alpha}\cdot \mathbf{p_{3\cdots n}} + i\epsilon) \cdots (\mathbf{e}_{\alpha}\cdot \mathbf{p_{n}} + i\epsilon)} \, .
    \label{eq:odintre}
\end{multline}
Where $\mathbf{p_{i\cdots m}} \equiv \mathbf{p_{i}} + \mathbf{p_{i+1}} + \cdots + \mathbf{p_{m-1}} + \mathbf{p_{m}}$. Substituting (\ref{eq:gdcop}) in (\ref{eq:bstar2}) and integrating with respect to $\alpha$, ${\tilde{B}}_a^\star (x^+;\mathbf{P})$ in momentum space reads 
\begin{multline}
    {\tilde{B}}_a^\star (x^+;\mathbf{P})
= {\tilde A}^\star_{a} (x^+;\mathbf{P}) + (-g)\int_{-\infty}^{+\infty}\! d^3\,\mathbf{p_1} d^3\, \mathbf{p_2}\frac{\delta^3 (\mathbf{P} - \mathbf{p_{12}})}{{\tilde v}^{\ast}_{1(12)}} \left\{ {\tilde A}^\bullet_{b_1} (x^+;\mathbf{p_1})\right.  \\
\left. \left(\frac{p_2^+}{p_{12}^+}\right )^2{\tilde A}^\star_{b_2} (x^+;\mathbf{p_2}) +  \left(\frac{p_1^+}{p_{12}^+}\right )^2  {\tilde A}^\star_{b_1} (x^+;\mathbf{p_1})  {\tilde A}^\bullet_{b_2} (x^+;\mathbf{p_2}) \right\} \mathrm{Tr}(t^a t^{b_1} t^{b_2}) \\
+ (-g)^2 \int_{-\infty}^{+\infty}\! d^3\,\mathbf{p_1} d^3\, \mathbf{p_2} d^3\, \mathbf{p_3} \frac{\delta^3 (\mathbf{P} - \mathbf{p_{123}})}{\Tilde{v}^*_{1(123)}\Tilde{v}^*_{(12)(123)}}   \left\{{\tilde A}^\bullet_{b_1} (x^+;\mathbf{p_1}){\tilde A}^\bullet_{b_2} (x^+;\mathbf{p_2})\right.  \\
\left. \left(\frac{p_3^+}{p_{123}^+}\right )^2{\tilde A}^\star_{b_3} (x^+;\mathbf{p_3}) +  {\tilde A}^\bullet_{b_1} (x^+;\mathbf{p_1})   \left(\frac{p_2^+}{p_{123}^+}\right )^2{\tilde A}^\star_{b_2} (x^+;\mathbf{p_2}){\tilde A}^\bullet_{b_3} (x^+;\mathbf{p_3}) \right. \\
 \left.  + \left(\frac{p_1^+}{p_{123}^+}\right )^2{\tilde A}^\star_{b_1} (x^+;\mathbf{p_1}){\tilde A}^\bullet_{b_2} (x^+;\mathbf{p_2}){\tilde A}^\bullet_{b_3} (x^+;\mathbf{p_3}) \right\}\mathrm{Tr}(t^a t^{b_1} t^{b_2} t^{b_3})+ \cdots 
\label{eq:bstarmom}
\end{multline}
An important point to note, which will be useful later, is that the kernels in (\ref{eq:bstarmom}) (outside the curly braces) are exactly same as the Wilson line (\ref{eq:WilsonLineSol}) coefficient functions in momentum space (\ref{eq:Gamma_n}) denoted by $\tilde{\Gamma}_n (\mathbf{P};\mathbf{p}_1,\dots,\mathbf{p}_n)$. Each term in (\ref{eq:bstarmom}) (except for the first term which is unity) can be compactly written in terms of $\tilde{\Gamma}_n (\mathbf{P};\mathbf{p}_1,\dots,\mathbf{p}_n)$.
The second term of (\ref{eq:bstarmom}), following the notations (\ref{eq:psiperm1}),  (\ref{eq:psiperm2}) reads
\begin{multline}
    \left[{\tilde{B}}_a^\star (x^+;\mathbf{P})\right]_{2nd} = 2 \int_{-\infty}^{+\infty}\! d^3\,\mathbf{p_1}\, d^3\, \mathbf{p_2}\, \Gamma_2(12)\mathrm{Tr}(a12)\\
    \left\{ {\tilde A}^\bullet_{b_1} (x^+;\mathbf{p_1})\left(\frac{p_2^+}{p_{12}^+}\right )^2{\tilde A}^\star_{b_2} (x^+;\mathbf{p_2}) +     {\tilde A}^\star_{b_1} (x^+;\mathbf{p_1}) \left(\frac{p_1^+}{p_{12}^+}\right )^2 {\tilde A}^\bullet_{b_2} (x^+;\mathbf{p_2}) \right\} \, .
\end{multline}
Above, renaming $(\mathbf{p_1}, b_1) \leftrightarrow (\mathbf{p_2}, b_2)$ in the first term we get
\begin{multline}
    \left[{\tilde{B}}_a^\star (x^+;\mathbf{P})\right]_{2nd} = 2 \int_{-\infty}^{+\infty}\! d^3\,\mathbf{p_1} d^3\,  \mathbf{p_2}\,{\tilde A}^\star_{b_1} (x^+;\mathbf{p_1}){\tilde A}^\bullet_{b_2} (x^+;\mathbf{p_2}) \\
\times \left(\frac{p_1^+}{p_{12}^+}\right )^2  \left[\Gamma_2(12)\mathrm{Tr}(a12) + \Gamma_2(21)\mathrm{Tr}(a21)    \right] \, .
\end{multline}
Compactly we have
\begin{multline}
     \left[{\tilde{B}}_a^\star (x^+;\mathbf{P})\right]_{2nd} = \int_{-\infty}^{+\infty}\! d^3\,\mathbf{p_1}\, d^3\, \mathbf{p_2}\,{\tilde A}^\star_{b_1} (x^+;\mathbf{p_1}){\tilde A}^\bullet_{b_2} (x^+;\mathbf{p_2}) \\
    \left[ 2  \left(\frac{p_1^+}{p_{12}^+}\right )^2 {\tilde \Gamma}_2^{a b_1 b_2}(\mathbf{P}; \mathbf{p_{1}}, \mathbf{p_{2}} )     \right] \, .
    \label{eq:bsttwo}
\end{multline}
 We see that the expression in the square bracket is exactly $\tilde{\Upsilon}_2$ of (\ref{eq:Upsilon_n}).

The third term of (\ref{eq:bstarmom}) is
\begin{multline}
    \left[{\tilde{B}}_a^\star (x^+;\mathbf{P})\right]_{3rd} = (3!) \int_{-\infty}^{+\infty}\! d^3\,\mathbf{p_1} \,d^3\, \mathbf{p_2} \,d^3\, \mathbf{p_3} \,\Gamma_3 (123) \mathrm{Tr} (a123)\left\{{\tilde A}^\bullet_{b_1} (x^+;\mathbf{p_1})\right.  \\
 \left.{\tilde A}^\bullet_{b_2} (x^+;\mathbf{p_2}) \left(\frac{p_3^+}{p_{123}^+}\right )^2{\tilde A}^\star_{b_3} (x^+;\mathbf{p_3})+  {\tilde A}^\bullet_{b_1} (x^+;\mathbf{p_1})   \left(\frac{p_2^+}{p_{123}^+}\right )^2{\tilde A}^\star_{b_2} (x^+;\mathbf{p_2})\right.  \\
 \left. {\tilde A}^\bullet_{b_3} (x^+;\mathbf{p_3})  + \left(\frac{p_1^+}{p_{123}^+}\right )^2{\tilde A}^\star_{b_1} (x^+;\mathbf{p_1}){\tilde A}^\bullet_{b_2} (x^+;\mathbf{p_2}){\tilde A}^\bullet_{b_3} (x^+;\mathbf{p_3}) \right\} \, .
\end{multline}
Above, renaming the variables such that we have ${\tilde A}^\star_{b_1} (x^+;\mathbf{p_1})$ in all the terms, we get
\begin{multline}
    \left[{\tilde{B}}_a^\star (x^+;\mathbf{P})\right]_{3rd} = (3!) \int_{-\infty}^{+\infty}\!d^3\, \mathbf{p_1} d^3\, \mathbf{p_2} d^3\, \mathbf{p_3}\,{\tilde A}^\star_{b_1} (x^+;\mathbf{p_1}){\tilde A}^\bullet_{b_2} (x^+;\mathbf{p_2}) \\
 {\tilde A}^\bullet_{b_3} (x^+;\mathbf{p_3}) \left(\frac{p_1^+}{p_{123}^+}\right )^2 \left[ \Gamma_3 (123) \mathrm{Tr} (a123) + \Gamma_3 (213) \mathrm{Tr} (a213)\right.  \\
 \left. +\Gamma_3 (321) \mathrm{Tr} (a321)     \right] \, .
\end{multline}
Exploring the symmetry with respect to the pairs $(b_2, \mathbf{p_{2}})$ and $(b_3, \mathbf{p_{3}})$, keeping $(b_1, \mathbf{p_{1}})$ fixed, we rewrite the above expression as
\begin{multline}
    \left[{\tilde{B}}_a^\star (x^+;\mathbf{P})\right]_{3rd} = (3!) \int_{-\infty}^{+\infty}\! d^3\,\mathbf{p_1} d^3\, \mathbf{p_2} d^3\, \mathbf{p_3}\,{\tilde A}^\star_{b_1} (x^+;\mathbf{p_1}){\tilde A}^\bullet_{b_2} (x^+;\mathbf{p_2})\\
{\tilde A}^\bullet_{b_3} (x^+;\mathbf{p_3}) \left(\frac{p_1^+}{p_{123}^+}\right )^2
  \left[ \frac{1}{2}\Gamma_3 (123) \mathrm{Tr} (a123) +\frac{1}{2}\Gamma_3 (132) \mathrm{Tr} (a132) \right.  \\
+ \left. \frac{1}{2}\Gamma_3 (213) \mathrm{Tr} (a213)+ \frac{1}{2}\Gamma_3 (312) \mathrm{Tr} (a312) \right.  \\
+\left. \frac{1}{2}\Gamma_3 (321) \mathrm{Tr} (a321) +\frac{1}{2}\Gamma_3 (231) \mathrm{Tr} (a231)    \right] \, .
\end{multline}
Compactly,
\begin{multline}
     \left[{\tilde{B}}_a^\star (x^+;\mathbf{P})\right]_{3rd} = \int_{-\infty}^{+\infty}\! d^3\,\mathbf{p_1} d^3\, \mathbf{p_2} d^3\, \mathbf{p_3}\,{\tilde A}^\star_{b_1} (x^+;\mathbf{p_1}){\tilde A}^\bullet_{b_2} (x^+;\mathbf{p_2})\\
    {\tilde A}^\bullet_{b_3} (x^+;\mathbf{p_3})\left[ 3  \left(\frac{p_1^+}{p_{123}^+}\right )^2 {\tilde \Gamma}_3^{a b_1 b_2 b_3}(\mathbf{P}; \mathbf{p_{1}}, \mathbf{p_{2}}, \mathbf{p_{3}} )     \right]\,.
    \label{eq:bstth}
\end{multline}
Comparing with (\ref{eq:Upsilon_n}) we see that the expression in the square bracket is $\tilde{\Upsilon}_3$.

The $n^{th}$ term of (\ref{eq:bstarmom}) is
\begin{multline}
    \left[{\tilde{B}}_a^\star (x^+;\mathbf{P})\right]_{n^{th}} = n! \int_{-\infty}^{+\infty}\! d^3\, \mathbf{p_1} \cdots d^3 \mathbf{p_n}\,  \Gamma_n (12\cdots n) \mathrm{Tr}  (a12\cdots n)\\
\left\{{\tilde A}^\bullet_{b_1} (x^+;\mathbf{p_1})\cdots{\tilde A}^\bullet_{b_{n-1}} (x^+;\mathbf{p_{n-1}}) \left(\frac{p_n^+}{p_{1\cdots n}^+}\right )^2{\tilde A}^\star_{b_n} (x^+;\mathbf{p_n})  \right.  \\
 \left. +  {\tilde A}^\bullet_{b_1} (x^+;\mathbf{p_1}) \cdots  \left(\frac{p_{n-1}^+}{p_{1\cdots n}^+}\right )^2{\tilde A}^\star_{b_{n-1}} (x^+;\mathbf{p_{n-1}}){\tilde A}^\bullet_{b_n} (x^+;\mathbf{p_n}) \right.  \\
 \left.+ \cdots + \left(\frac{p_1^+}{p_{1\cdots n}^+}\right )^2{\tilde A}^\star_{1_n} (x^+;\mathbf{p_1}){\tilde A}^\bullet_{b_2} (x^+;\mathbf{p_2})\cdots{\tilde A}^\bullet_{b_n} (x^+;\mathbf{p_n}) \right\} \, .
 \label{eq:bnexp}
\end{multline}
We introduce the following notation for convenience
\begin{eqnarray}
\left(\frac{p_i^+}{p_{1\cdots n}^+}\right )^2 &=& \Lambda_i    \label{lamdef}\\
{\tilde A}^\star_{b_i} (x^+;\mathbf{p_i}) &=& {\tilde A}^\star_i    \hspace{1.5cm} {\tilde A}^\bullet_{b_j} (x^+;\mathbf{p_j}) = {\tilde A}^\bullet_j  \label{asbn}
\end{eqnarray}
Renaming each of the $n$ terms  in (\ref{eq:bnexp}) such that we have ${\tilde A}^\star_{b_1} (x^+;\mathbf{p_1})$ and $ \Lambda_1$ in all of them, we get 
\begin{multline}
    \left[{\tilde{B}}_a^\star (x^+;\mathbf{P})\right]_{n^{th}} = n! \int_{-\infty}^{+\infty}\! d^3\,\mathbf{p_1} \cdots d^3\, \mathbf{p_n}\, {\tilde A}^\star_{1}\,{\tilde A}^\bullet_{2} \cdots{\tilde A}^\bullet_{n}\,  \Lambda_1\\
\bigg[ \Gamma_n (1\,2\cdots n)\, \mathrm{Tr}  (a\,1\,2\cdots n)\, + \,\Gamma_n (2\,1\,3\cdots n)\, \mathrm{Tr}  (a\,2\,1\,3\cdots n) +\cdots  \\
   + \Gamma_n ((n-1)\, 2 \cdots(n-2) 1\, n)\,\mathrm{Tr}  (a\,(n-1)\, 2 \cdots(n-2)\, 1\, n)  \\
  + \Gamma_n (n\, 2 \cdots(n-1)\, 1) \, \mathrm{Tr}  (a\, n\, 2 \cdots(n-1)\, 1)     \bigg]\, .
\end{multline}
The above expression has $n$ terms in the integral. Each term has symmetry with respect to the pairs $(b_n, \mathbf{p_{n}}) \cdots (b_2, \mathbf{p_{2}})$. Thus keeping $(b_1, \mathbf{p_{1}})$ fixed, each term can be written as a sum of $(n-1)!$ terms (all permutation of $2\cdots n$). With this we get
\begin{multline}
     \left[{\tilde{B}}_a^\star (x^+;\mathbf{P})\right]_{n^{th}} = (n!) \int_{-\infty}^{+\infty}\! d^3\,\mathbf{p_1} \cdots d^3\, \mathbf{p_n}\, {\tilde A}^\star_{1}\,{\tilde A}^\bullet_{2} \cdots{\tilde A}^\bullet_{n} \, \Lambda_1 \frac{1}{(n-1)!}\\
    \sum_{\mathrm{permutations}} \mathrm{Tr} (a 1 \cdots n)\Gamma_n (1 \cdots n)\, .
\end{multline}
Compactly we have
\begin{multline}
     \left[{\tilde{B}}_a^\star (x^+;\mathbf{P})\right]_{n^{th}} = \int_{-\infty}^{+\infty}\! d^3\,\mathbf{p_1} \cdots d^3 \mathbf{p_n} \,{\tilde A}^\star_{b_1} (x^+;\mathbf{p_1}) {\tilde A}^\bullet_{b_2} (x^+;\mathbf{p_2})
    \cdots\\
    {\tilde A}^\bullet_{b_n} (x^+;\mathbf{p_n})  \left[ n\left(\frac{p_1^+}{p_{1\cdots n}^+}\right )^2 {\tilde \Gamma}_{n}^{a b_1 \cdots b_n }(\mathbf{P}; \mathbf{p_{1}}  \cdots \mathbf{p_{n}} )    \right] \, .
    \label{eq:bstn}
\end{multline}
The expression in square bracket is exactly (\ref{eq:Upsilon_n}). Thus we have proven that the position space functional (\ref{eq:Bstar_WL1}) corresponds to (\ref{eq:Bstar}) with (\ref{eq:Upsilon_n}), upon transforming to momentum space.

\end{document}